\documentclass[journal,twoside,web]{ieeecolor}
\usepackage{generic}
\usepackage{cite}
\usepackage{amsmath,amssymb,amsfonts}
\usepackage{algorithmic}
\usepackage{graphicx}
\usepackage{textcomp}
\usepackage[normalem]{ulem}  

\usepackage{subcaption} 
\usepackage{url}

\usepackage{tikz}
\newcommand\copyrighttext{%
  \normalsize This work has been submitted to the IEEE for possible publication. Copyright may be transferred without notice, after which this version may no longer be accessible. 
  }
\newcommand\copyrightnotice{%
\begin{tikzpicture}[remember picture,overlay]
\node[anchor=south] at (current page.south) {\fbox{\parbox{\dimexpr\textwidth-\fboxsep-\fboxrule\relax}{\copyrighttext}}};
\end{tikzpicture}%
}

\def\x#1{} 
\def\liml{\mathop{\lim}\limits} 
\def\maxl{\mathop{\max}\limits} 
\def\suml{\mathop{\sum}\limits} 
\def\cdc{,\ldots,}
\def\N{{\mathbb N}}
\def\R{{\mathbb R}}
\def\C{{\mathbb C}}
\def\VV{\mathcal{V}}
\def\EE{\mathcal{E}}
\def\GG{\mathcal{G}}
\def\AA{\mathcal{A}}

\def\LL{\mathcal{L}}
\def\tY{{\widetilde Y}}
\def\la{\lambda}

\newtheorem{remark}{Remark}
\newtheorem{theorem}{Theorem}
\newtheorem{lemma}{Lemma}
\newtheorem{definition}{Definition}
\newtheorem{corollary}{Corollary}
\newtheorem{example}{Example}

\newcommand{\e}{\mathop{\mathrm{e}}}

\definecolor{ggreen}{rgb}{0,.6,0}
\newcommand{\fich}[1]{{\color{black}#1}} 
\def\pars#1{{\color{black}#1}} 
\def\ch#1{{\color{black}#1}} 
\def\chh#1{{\color{black}#1}} 
\def\chp#1{{\color{black}#1}}
\def\chq#1{{\color{black}#1}}    
\def\chno#1{{\ch{\sout{#1}}}}
\def\shch#1{{\color{black}#1}} 

\makeatletter
\renewcommand*\env@matrix[1][c]{\hskip -\arraycolsep
	\let\@ifnextchar\new@ifnextchar
	\array{*\c@MaxMatrixCols #1}}
\makeatother

\def\BibTeX{{\rm B\kern-.05em{\sc i\kern-.025em b}\kern-.08em
    T\kern-.1667em\lower.7ex\hbox{E}\kern-.125emX}}
\begin{document}
\title{Hierarchical Cyclic Pursuit: Algebraic Curves Containing the Laplacian Spectra}
\author{Sergei E. Parsegov, Pavel Yu. Chebotarev, Pavel S. Shcherbakov, Federico M. Ib\'a\~nez
\thanks{S.~Parsegov and F.~Ib\'a\~nez are with the Center for Energy Science and Technology, Skolkovo Institute of Science and Technology, Moscow, 121205, Russia (e-mail: s.parsegov@skoltech.ru; fm.ibanez@skoltech.ru).}
\thanks{P.~Chebotarev is with Moscow Institute of Physics and Technology, 9~Institutskiy per., Dolgoprudny, Moscow Region, 141701, Russia (e-mail: pavel4e@gmail.com).}
\thanks{P.~Shcherbakov is with the Institute of Control Sciences, Russian Academy of Sciences, Moscow, 117997, Russia, and Moscow Institute of Physics and Technology, 9, Insitute lane, Dolgoprudny 141701, Russia (e-mail: cavour118@mail.ru).}
}

\copyrightnotice
\maketitle

\begin{abstract}
The paper addresses the problem of multi-agent communication in networks with regular \ch{directed} ring structure. \ch{These} can be viewed as hierarchical extension\ch{s} of \ch{the} classical cyclic pursuit topology. We show that the spectra of the corresponding Laplacian matrices \ch{allow} \pars{exact} localization on the complex plane.
\ch{Furthermore}, we derive a general form of the characteristic polynomial \ch{of such matrices}, analyze the algebraic curves its roots belong to, and propose a way to obtain their closed-form equations. In combination with\x{a} frequency domain consensus \ch{criteria} for high-order SISO linear agents, these curves \ch{enable one to analyze the} feasibility of consensus in networks with \ch{varying} number of agents.

\end{abstract}


\section{Introduction}
\label{sec:introduction}

\ch{The Laplacian spectra of graphs} play \ch{an} important role in \ch{solving} distributed optimization and control problems, since they \ch{mainly determine the} stability and the convergence rate \ch{of the corresponding dynamical systems}~\cite{Bullo2018,Rogozinetal2020,CheAga2009}
For a fixed graph, \chh{finding the spectrum does not cause any difficulties}\x{this is easily doable}, but if we consider graphs with a scalable structure (i.e., those constructed by \ch{the} repetition of the same component), the problem of exact \chh{calculation or localization of the spectrum turns out to be non-trivial}.
A huge amount of literature is devoted to the derivation of formulas for the Laplacian spectra of \emph{undirected topologies}, including various lattices such as rectangular grids, honeycombs (see \cite{Pozrikidis2014} and the references therein), hierarchical small-world networks \cite{Liuetal2015}, products and coronas of graphs \cite{Kammerdineretal2017}, and many others.

However, \ch{when analyzing the} dynamics of network systems, \emph{directed} communication topologies are of \ch{major interest}. \x{For example, it may happen}Say, it can be observed that a group of \ch{high-order} agents \ch{may converge to consensus} under \ch{an} undirected interaction topology, \ch{however, it fails to do so under the corresponding uni-directed one, even though this topology contains} a spanning converging tree. \ch{A precise} localization of \ch{the} Laplacian spectra of digraphs \chh{serves as the basis for the analysis} of consensus \ch{problems in such situations}.

In \chh{this paper}\x{what follows}, we study \ch{several} generalizations of the \emph{cyclic pursuit} multi-agent strategy. Its history can be traced back to 1878, when J.G.~Darboux published his elegant work \cite{Darboux1878}, where \chh{he}\x{the author} studied \chh{some} geometric averaging \ch{procedure} and proved its convergence to consensus. Basically, cyclic pursuit is a strategy where agent $i$ pursues its neighbor $i - 1$ modulo $N$, where $N$ is \ch{the} number of agents. Evidently, such a communication structure is a uni-directed ring \fich{or a ``predecessor–follower'' topology}, \ch{i.e.,} a Hamiltonian cycle.

Cyclic pursuit strategies attracted the attention of different scientific communities (see, e.g., \cite{Brucksteinetal1991, Bruckstein1993, Nahin2007, Sharmaetal2013, ElorBruckstein2011, MarshallBrouckeFrancis2004} and references therein) \ch{due to} a wide range of applications including but not limited to numerous formation control tasks \ch{such} as patrolling, boundary mapping, etc.  \ch{Their} extensions to hierarchical structures were considered in \cite{SmithBrouckeFrancis2005, DingYanLin2009, MukherjeeGhose2016, Parsegovetal2022}; papers \cite{SinhaGhose2006}, \cite{Mukherjee2021} addressed the case of heterogeneous agents; the effect of communication delays was analyzed in~\cite{Deetal2020}; geometrical problems related to cyclic pursuit-like algorithms \ch{were} \chh{studied} in \cite{ElmachtoubVanLoan2010} and~\cite{Shcherbakov2011}. Some pursuit algorithms \ch{use} the rotation operator in order to follow \ch{desired} trajectories, see \cite{RamirezRiberosPavoneetal2010} and the references therein.
\textcolor{black}{The paper~\cite{AgaChe2012} shows the connection of discrete-time weighted cyclic pursuit with \chp{the}\x{a} general DeGroot model.}
Another group of strategies
(protocols) \ch{is} based on bi-directional topologies\chh{\cite{MukherjeeZelazo2018}}, \ch{that is}, each agent $i$ has \ch{a relative} information \ch{about its} neighbors $i-1$ and $i+1$ (modulo $N$\x{with $0 \equiv N$}).\x{\ch{In this manner, the agents in} \cite{MukherjeeZelazo2018} are connected by a bi-directional ring.} The row straightening\x{OK!} problems studied in\chh{\cite{WagnerBruckstein1997,KvintoParsegov2012,ProskurnikovParsegov2016}} also imply symmetric communications except for fixed ``anchors'' (the endpoints of \ch{a} segment). \fich{The problems of {\it vehicle platooning} with cyclic communications (see, e.g., \cite{RoggeAeyels2008, Piranietal2022, Stuedlietal2018, Hermanetal2013}) are also closely related to the problems of cyclic pursuit. In this case, the network system also has inputs including the desired inter-vehicular distances and communication disturbances. The analysis of the closed-loop stability of such systems is reduced to the study of state matrices close or identical to those studied in cyclic pursuit.}

\ch{Regular ring structures model symmetric hierarchical \chh{interaction}\x{communications} between agents.\x{they are theoretically attractive as} In some cases, \chh{these structures allow for}\x{one to obtain} closed-form expressions for the spectra of the corresponding Laplacian matrices, which \chh{helps to analyse the}\x{simplifies the analysis of} control protocols these matrices are involved in.}

\ch{While} cyclic pursuit can be treated\x{considered} as a special case of consensus \ch{seeking}, the properties of the underlying interaction topology \ch{are closely related to \chh{classical}\x{certain} mathematical considerations including the\x{classical} study of algebraic curves. For the basic cyclic pursuit topology,} the eigenvalues of the corresponding Laplacian matrix are roots of unity \cite{SmithBrouckeFrancis2005}\x{, and thus \ch{they} can be found in closed form. Moreover,}: no matter how many agents/nodes constitute the network, the spectrum lies on \ch{the} unit circle. This fact \ch{prompted} us to study \ch{hierarchical and other} \chh{generalized} ring topologies, \ch{which \chh{led}\x{lead} to higher-order curves} \fich{that contain their Laplacian spectra}. 

In \chh{this paper}\x{particular}, we study ring digraphs with a \fich{hierarchical} ``necklace'' 
structure. \fich{It is convenient to \chq{explore the Laplacian spectra}\x{study the spectral properties of the Laplacian matrices} of such\x{ ring} graphs with regularly interleaved directed and undirected arcs\x{,} using the concept of hierarchy. \x{In the paper, for this}\chq{Namely}, we introduce a {\it macro-vertex}, \chq{which is}\x{ represented by} a sequence of directed and undirected arcs (the lower level of the hierarchy) and a {\it directed ring \chq{of macro-vertices}} (the upper level of the hierarchy). The topologies constructed in this way \chq{occupy} an intermediate position between directed and undirected rings\chq{, which have been} widely studied in relation to cyclic pursuit and control of homogeneous vehicular platoons running on a ring \chq{(}see, e.g., the nearest neighbor ring topologies presented in Fig.~2~(h) and~(i) in \cite{Piranietal2022}).}



\fich{
A \x{convenient}\chq{useful} classification of consensus problems \chq{based on} the notion of\x{ a} {\it complexity space} was proposed in \cite[\x{see}Fig.~1.1]{Wieland2010}. In accordance with it, three independent dimensions of complexity can be identified in which the simplest first-order consensus model can be generalized, namely, (1)~the complexity of the agent model, (2)~topological complexity (complexity of the structure of interactions), \chq{and} (3)~the complexity of couplings between agents.
\x{Taking it into account,}\chq{The} contribution of our paper to the general study of consensus in network systems can be attributed to \chq{the} first two directions: \x{1)~}the analysis and localization of the Laplacian spectra of special ring topologies \chq{to~(2)} and \x{2)~}complex high-order models of agents \chq{to~(1)}. Specifically, we prove that the Laplacian spectra of the studied\x{ ring} digraphs lie on certain high-order algebraic curves\x{ no matter how many} \chq{ irrespective of the number of macro-vertices forming} the network.
Along with this, we present an algorithm for obtaining equations of these curves. Based on this localization, we propose a geometric consensus condition in the frequency domain applicable to any number of interacting agents.}

The paper is organized as follows. \ch{Section~\ref{sec_prelim}} \x{presents}introduces some mathematical preliminaries needed for the \chh{subsequent analysis}\x{exposition to follow,} and discusses the statement of the problem. 
\pars{The main results\x{of the paper} \chh{that describe} the Laplacian spectra of ring digraphs \ch{are presented in Section~\ref{sec:LapSpe}}.} \ch{We} prove that, regardless of the number of macro-vertices in \ch{such a digraph}, \ch{its} Laplacian spectrum lies on \ch{a certain} algebraic curve and \ch{provide an} algorithm to derive an \ch{implicit
equation (of the form $p(x,y)=0$) of this} curve \chh{in~$\R^2$}.
\ch{In Section~\ref{sec_consens}, we study \chh{consensus problems for} a group} of high-order linear SISO agents interacting through the discussed ring topologies, \ch{that is, performing} {\em hierarchical\/} cyclic pursuit. We apply the frequency domain criterion \cite{PolyakTsypkin1996, HaraTanakaIwasaki2014, LiDuan2017} to derive a \pars{necessary and sufficient} consensus condition, \ch{which} does not depend on the number of agents in the network\x{\pars{OR BETTER: for infinitely large number of agents in the network}}. The theoretical results are accompanied with numerical illustrations
and, finally,\x{the} conclusions are given.

Throughout the paper\ch{,} $j\!:=\!\sqrt {-1}$ denotes the imaginary unit, while letters $i$ and $k$ are used\x{as indices} \chh{for indexing purposes}.



\section{Preliminaries and Problem Statement}\label{sec_prelim}

In this paper, we study network systems \ch{that have} hierarchical ring structure. After defining\x{some} \chh{the} basic terminology, we formulate the problem.

Throughout the paper\ch{, we consider finite digraphs allowing in some cases multiple arcs and loops. A} \ch{di}graph is denoted by $\GG_N = (\VV,\EE)$, where
\pars{$\VV = \{1\cdc N\}$} stands for the node set and $\EE$ \chh{for}\x{is} the \ch{multi}set\footnote{\ch{A multiset, unlike a set, allows multiple occurrences of each element. We need this in one particular case in which we assume the presence of multiple arcs in a digraph (see Fig.~\ref{fig:sr-2}).}} of arcs.

The formal definitions of the adjacency and Laplacian matrices of \ch{an} unweighted digraph $\GG_N$ are given below.

\begin{definition}
	\label{def:adjacency}
	The {\em adjacency matrix\/}\x{$\AA_N$} associated with \ch{a digraph} $\GG_N = (\VV,\EE)$ is the matrix $\AA_N = (a_{ik})\in\R^{N \times N} $, where \ch{each entry $a_{ik}$ is the number of arcs {of the form} $(i,k)$ in} $\mathcal{E}.$
\end{definition}
\begin{definition}
	\label{def:laplacian}
	The {\em Laplacian matrix\/} $\LL_N \in \R^{N \times N}$ of\x{a digraph} $\GG_N$ is the matrix with entries $l_{ii} = \sum_{\ch{k\ne i}}a_{ik}$ and $l_{ik} = -a_{ik}$ for $i \neq k,$ \chh{where $(a_{ik})=\AA_N$ is the adjacency matrix of~$\GG_N$}.
\end{definition}

\begin{figure}[ht] 
	\centering
	\begin{subfigure}[b]{0.12\textwidth}
		\includegraphics[width=\textwidth]{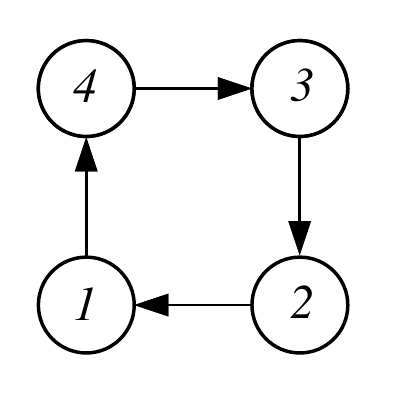}
		\caption{\ch{The~}Hamiltonian cycle \ch{on} 4 nodes}
		\label{fig:ham}
	\end{subfigure}
	~ 
	\begin{subfigure}[b]{0.12\textwidth}
		\includegraphics[width=\textwidth]{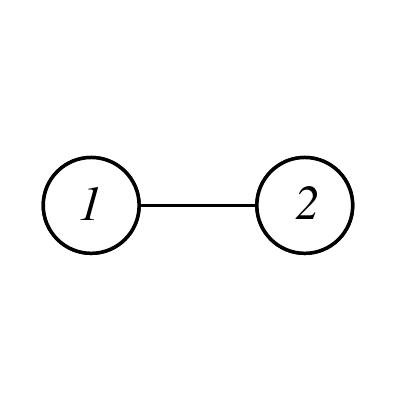}
		\caption{\ch{A~}macro-vertex on 2 nodes}
		\label{fig:macro2}
	\end{subfigure}
	\caption{A Hamiltonian cycle corresponding to the cyclic pursuit strategy with four agents (a) and a macro-vertex (b) \label{fig_pic_00}}
\end{figure}

For example, consider a graph that \chh{represents}\x{reflects} communications within \ch{the} conventional cyclic pursuit strategy \ch{\chh{for}\x{of} four agents} (Fig.~\ref{fig:ham}). \ch{Here, an arc from $i$ to $k$ shows that agent $i$ pursues agent~$k.$}
	
The corresponding Laplacian matrix \pars{for \ch{the} general case \chh{of}\x{with} $N$ agents can be defined through the \ch{counter-clockwise principal circulant permutation matrix \cite{Johnsen1971}} $\mathcal{P}_N$ as follows:
$$
\LL_N = I_N - \mathcal{P}_N,
$$
where $I_N \in \mathbb{R}^{N\times N}$ is \ch{the} identity 
matrix,
\begin{equation}
\label{eq:perm_cycl_purs}
\mathcal{P}_N =
\begin{bmatrix}[r]
0 & 0 & 0 & \cdots & 1\\
1 & 0 & 0 & \cdots & 0 \\
0 & 1 & 0 & \cdots & 0  \\
\vdots & \vdots & \ddots & \ddots & \vdots \\
0 & \cdots & 0 & 1 & 0
\end{bmatrix},
\end{equation}
and
\begin{equation}
\label{eq:laplacian_cycl_purs}
\LL_N =
\begin{bmatrix}[r]
1 & 0 & 0 &\cdots & -1\\
-1 & 1 & 0 & \cdots & 0 \\
0 & -1 & 1 & \cdots & 0  \\
\vdots & \vdots & \ddots & \ddots & \vdots \\
0 & \cdots & 0 & -1 & 1
\end{bmatrix}.
\end{equation}}

We now \chh{describe}\x{begin by} the structure of hierarchical network systems studied \ch{below}. The lower level of the hierarchy is a \ch{{\it linear macro-vertex}, \chh{which is}\x{or} a specific subdigraph whose $n\ge1$ nodes are identified with} indexed dynamical agents, \ch{while} the \chh{top} level is a Hamiltonian cycle on\footnote{\ch{The shortest Hamiltonian cycle consists of one node (in \chh{our construction}\x{the present case}, it is a macro-vertex) and one directed loop.}} \ch{$m\ge1$ macro-}vertices.

\begin{definition}\x{x}
\label{def:mv}
\ch{A {\em linear macro-vertex\/} \chh{$\GG^i_n = (\VV^i,\EE^i)$} of a digraph $\GG_N = (\VV,\EE)$ is a subdigraph of $\GG_N$ with \chh{$\VV^i=\{v^i_1\cdc v^i_n\}$ ($n\ge1$)} obtained from the directed path $v^i_n\to v^i_{n-1}\to\cdots\to v^i_1$ ({\em main direction}; no arcs when $n=1$) by adding the reverse path $v^i_1\to v^i_2\to\cdots\to v^i_n$ 
from which any subset of arcs is dropped.}
\end{definition}

\chh{The following definition introduces a topology consisting of $m$ identical macro-vertices on disjoint subsets of nodes along with a top-level Hamiltonian cycle that forms a Hamiltonian cycle on the whole set of $N=mn$ nodes together with the main direction paths traversing the macro-vertices.}%
\x{It is assumed that~\ch{(i)} all macro-vertices \ch{of a topology} are identical \ch{up to the enumeration of nodes and (ii) the upper level} Hamiltonian cycle \ch{together with the main direction paths of the macro-vertices form a Hamiltonian cycle on the cumulative set of $N=mn$ nodes.}}
We will associate the term {\em ring digraph} with such a topology.
\x{\ch{More formally this notion is introduced by the following definition.}}

\begin{definition}\x{x}
\label{def:ring_dg}
\ch{A {\it ring digraph} \chh{denoted by $\GG_{m,n}=(\VV,\EE)$} is a digraph such that 
$\VV=\bigcup_{i=1}^m\VV^i$, $\VV^i=\{n(i-1)+1\cdc ni\},$
$\EE=\big(\bigcup_{i=1}^m\EE^i\big)\cup\{e_1,\ldots,e_m\},$
\chh{$(\VV^i,\EE^i)=\GG^i_n$} are identical linear macro-vertices on $n\ge1$ nodes, and the arcs
$e_i=(ni+1,ni)$ $(i\in \pars{\{1\cdc m-1\}})$ and $e_m=(1,nm)$ link the first node of each macro-vertex with the $n$th node of the previous one (which is the same \chh{macro-vertex}\x{one} when $m=1$).}
\end{definition}

\x{x}
\ch{It can be observed that each macro-vertex of a ring digraph is its \emph{induced}%
\/\footnote{\ch{An induced subdigraph of a digraph is a subdigraph whose arc set consists of all of the arcs of the digraph that have both endpoints in the node set of the subdigraph.}} subdigraph whenever $m>1$, while for $m=1$, it drops the arc $(1,n).$ The arcs $e_1\cdc e_m$ form a Hamiltonian cycle on $m$ macro-vertices.}

\begin{figure}[ht]  
	\centerline{\includegraphics[width=6.375cm]{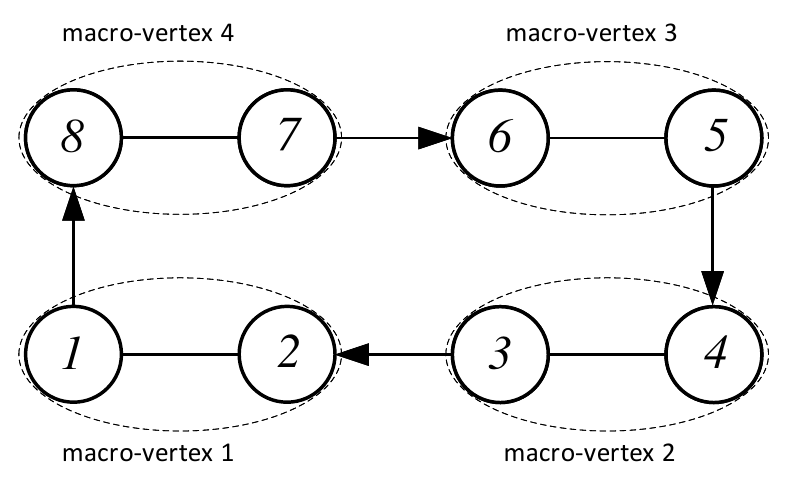}}
	\caption{\chh{A} ring digraph with four macro-vertices
	}
	\label{fig_pic_0000}
\end{figure}

An example of a \ch{ring digraph} with $n=2$ \ch{and} $m=4$ is \ch{presented} in Fig.~\ref{fig_pic_0000}. \ch{It is constructed from the Hamiltonian cycle shown in Fig.~\ref{fig:ham} and the macro-vertex (it is the complete digraph on} two nodes) \ch{shown in Fig.~\ref{fig:macro2}, where a pair of opposite arcs is represented by a line segment \chh{without arrows}}.

\begin{remark}\x{x}
\label{rem:ring_dg}
\ch{A ring digraph can be considered as} a Hamiltonian cycle  $\{(1,N), (N,N - 1), \ldots , (2, 1)\}$  supplemented \chh{by}\x{with} \ch{the path} $\{(1, 2), (2, 3)\cdc \ch{(N - 1,N)}\}$ in which $\nu$ $(\ch{0 \le \nu \le N-1})$
arcs are dropped in a regular fashion. In a sense, \ch{ring} digraphs fill the gap between \ch{\x{the}Hamiltonian cycle and\x{the} bi-directional} ring.
\ch{Obviously, every ring digraph contains a spanning converging tree.} \fich{It should be noted that this condition is necessary and sufficient for \chq{attaining asymptotic consensus in the system consisting of first-order agents.}\x{ the first-order consensus problem. Further in the text} \chq{In Section~\ref{sec_consens}, we consider}\x{, we discuss} a more general setting with high-order agent models and\x{ also} derive a consensus condition that does not depend on the number of nodes in the network.}
\end{remark}


\chh{We now} introduce cooperating agents and \ch{then} \ch{formulate} the problem. The agents are assumed to have identical high-order (double integrator \ch{or} higher) SISO linear models. \textcolor{black}{Let  $x_i \in \mathbb{R}$ represent the position of agent $i$, $i \in \{1\cdc N\}$.} Therefore, the \ch{consensus-seeking communication} \chh{over}\x{through} the network \ch{$\GG_{m,n}$} can be described as
\begin{eqnarray}
\label{eq:gen_cons_system0}
 & {\rm \bf a}(s)x_i = u_i, \\
 \label{eq:gen_cons_system01}
 & u_i = {\rm \bf b}(s)\left(\sum_{k \in \mathcal{N}_i}a_{ik}(x_k - x_i) \right), \quad i \in \pars{\{1\cdc N\}}\ch{,}
\end{eqnarray}
\pars{where $a_{ik}$  
\ch{are the elements of the adjacency matrix $\AA_N$ and}
$\mathcal{N}_i$ is \ch{the} set of neighbors of node~$i,$ \ch{i.e., the set of nodes~$k$ such that $a_{ik}\ne0$}.}
 Here $s :=  \frac{d}{dt}$ denotes the differentiation operator, the scalar polynomials
 $$
 {\rm \bf a}(s) = s^d+ {\rm \bf a}_{d-1}s^{d-1}+\ldots+{\rm \bf a}_{1}s +{\rm \bf a}_{0},
 $$
 $$
 {\rm \bf b}(s) = {\rm \bf b}_{q}s^q+ {\rm \bf b}_{q-1}s^{q-1}+\ldots+{\rm \bf b}_{1}s +{\rm \bf b}_{0}
 $$
 \ch{determine} 
 \ch{agent's} dynamics and communications, \ch{and $u_i$ is the control signal}.
For 
convenience, we assume $d>q$.

Let us introduce the vector $\xi_i = [x_i,~\dot x_i,\ldots, x^{(d-1)}_i]^{\top}$ and transform equations \eqref{eq:gen_cons_system0}, \eqref{eq:gen_cons_system01} into \ch{the} state-space \chh{form}\x{representation}
\begin{eqnarray}
\label{eq:gen_cons_system002}
\dot \xi_i  &=& A \xi_i + B u_i, \\
 \label{eq:gen_cons_system0003}
u_i &=& K\sum_{k \in \mathcal{N}_i}a_{ik}(\xi_k - \xi_i), \quad i \in \pars{\{1\cdc N\}},
\end{eqnarray}
where
$$
A = \begin{bmatrix}[r]
0 & 1 & 0 & \cdots & 0\\
0 & 0 & 1 & \cdots & 0\\
0 & 0 & 0 & \cdots & 0  \\
\vdots & \vdots & \vdots & \cdots & \vdots \\
0 & 0 & 0 & \cdots & 1 \\
-{\rm \bf a}_0 & -{\rm \bf a}_1 & -{\rm \bf a}_2 & \cdots & -{\rm \bf a}_{d-1}
\end{bmatrix}, \quad B = \begin{bmatrix}[r]
0 \\
0 \\
0 \\
\vdots \\
0 \\
1
\end{bmatrix},
$$
$$
K = \begin{bmatrix}[r] {\rm \bf b}_0 & {\rm \bf b}_1 & {\rm \bf b}_2 & \ldots & {\rm \bf b}_q & 0 & \ldots & 0  \end{bmatrix}.
$$

The entire closed-loop dynamics can thus be written as
\begin{equation}
    \label{eq:entire_consensus}
    \dot \xi = ({I}_N\otimes A-\LL_N\otimes {BK})\xi,
\end{equation}
where $\xi = [\xi^{\top}_1,~\xi^{\top}_2,\ldots, \xi^{\top}_{N}]^{\top}$\x{,} \ch{and $\otimes$ is the Kronecker product}.

In \chh{Section~\ref{sec_consens}}\x{this study}, we \chh{will} \ch{obtain} a consensus criterion for ring-shaped networks of agents \eqref{eq:gen_cons_system0}, \eqref{eq:gen_cons_system01}.

\chh{Let us formulate a} definition of consensus \chh{for the systems under study}.
\begin{definition}
\label{def:consen_def}
We say that the network system \eqref{eq:gen_cons_system002} with
feedback control \eqref{eq:gen_cons_system0003} {\em reaches consensus\/} if
\begin{equation}
    \label{eq:consen_def}
    \lim_{t \to \infty}{\| \xi_i(t) - \xi_k(t)\|} = 0, \quad \forall i,k \in \pars{\{1\cdc N\}}\x{,}
\end{equation}
for any initial condition $\xi(0) = [\xi_1^{\top}(0),\ldots, \xi_{N}^{\top}(0)]^{\top}$.
\end{definition}

In the simplest case \ch{of} ${\rm \bf a}(s) = s$ and ${\rm \bf b}(s) = 1$ we face \chh{the}\x{a} classical first-order consensus model; \pars{e.g., the cyclic pursuit if $a_{ik}=1$ for $k=i-1 \pmod N$} \chh{and $a_{ik}=0$ otherwise}. The \ch{corresponding} Laplacian matrix $\LL_N$ is given by \eqref{eq:laplacian_cycl_purs}, and its characteristic polynomial $\Delta(\la)$ has the form
$$
\Delta(\la) = (\la-1)^N-1.
$$

\chh{The} roots \chh{of $\Delta(\la)$} can be found using \ch{Lemma~\ref{lem.cyc}, which follows from De Moivre’s Theorem}.

\begin{lemma}
\label{lem.cyc}
The roots of the cyclotomic equation
\begin{equation}
\label{eq:cyclotomic1}
\sigma^N -1 = 0
\end{equation}
are
\begin{equation}
\label{eq:cyclotomic_roots1}
\sigma_k = {\rm e}^{j \frac{2 \pi k}{N}}, \; k \in \pars{\{0\cdc N-1\}},
\end{equation}
and the roots of
\begin{equation}
\label{eq:cyclotomic2}
\sigma^N + 1 = 0
\end{equation}
are
\begin{equation}
\label{eq:cyclotomic_roots2}
\sigma_k = {\rm e}^{j \frac{2 \pi k +\pi}{N}}, \; k \in \pars{\{0\cdc N-1\}}.
\end{equation}
The roots \ch{in} both sets are uniformly distributed on the unit circle \chh{centered at $(0,~j0)$} in the complex plane~\ch{$\C$}.
\end{lemma}

Therefore, the spectra of the Laplacian matrices \eqref{eq:laplacian_cycl_purs} \chh{with all $N\in\N$ are jointly dense on} the unit circle \ch{centered} at $(1,~j0)$.

\pars{\x{Evidently, }The \chh{equation of the} corresponding unit circle in $\mathbb{R}^2$ is\x{defined by}
\begin{equation}
    \label{eq:uncirc}
    (x-1)^2+y^2 - 1 = 0.
\end{equation}}
\ch{This} circle is \chh{a basic}\x{simplest} example of \ch{a curve} that contains the Laplacian \ch{spectrum of a} ring digraph; it entirely 
lies in \chh{$\C^+\cup \{0\}.$} \ch{The spectrum of any such a digraph contains $0$ with multiplicity~$1,$ which guarantees consensus in the \chh{first order} cyclic pursuit process according to the well-known consensus criterion}.
\fich{\begin{remark}
\label{rm:platoons}
The dynamic system \eqref{eq:gen_cons_system0}, \eqref{eq:gen_cons_system01} can be considered from different points of view: Its coordinates can have different physical meanings, and the signal $u_i$ can contain both the plant dynamics and elements of a local or/and a distributed controller. In addition, the right\chq{-}hand side can also contain additional external signals and disturbances that do not affect the form of the state matrix of the closed loop system \eqref{eq:entire_consensus}. A particular example of such a system is a leaderless vehicle platoon moving on a ring, see, e.g., \cite{RoggeAeyels2008, Piranietal2022, Stuedlietal2018, Hermanetal2013}. In such problems, two types of stability are studied: The classical stability of a closed loop system and string stability associated with the amplification of a disturbance propagating through the system (see \cite{Stuedlietal2017}, \cite{Monteiletal2019} and references therein). With an increase in the number of vehicles $N$ in the platoon, the system may exhibit {\it eventual instability} \cite{Stuedlietal2017}.
Therefore, the problem of stabilization regardless of the number $N$ is important.
\end{remark}}

The paper aims at:
\begin{itemize}
    \item \ch{localizing 
    the Laplacian spectra 
    of the ring digraphs defined above};
    \item obtaining a necessary and sufficient consensus condition
    \ch{applicable to any} number of agents in the network.
\end{itemize}

\section{Laplacian Spectra of Ring Digraphs}\label{sec:LapSpe}

\pars{In this \chh{section}\x{paper}, we propose a method for exact localization of Laplacian spectra for ring digraphs.
It \ch{turns out} that these spectra always lie on algebraic curves \chh{whose}\x{which} expressions \chh{can be found in closed form}.}
Thus, \pars{\x{closed-form expressions for} equations of \chh{these} curves\x{that contain the locus of the Laplacian spectra of ring digraphs}} \chh{are among} the main \ch{results} of the  work\x{work}\x{paper}.

\chh{First}\x{In this section}\ch{, we classify \chh{ring} digraphs\x{of this kind} and discuss their properties}. After that we
\begin{itemize}
\item derive a general form of the characteristic polynomial of \ch{the} corresponding Laplacian matrices;
\item \ch{present a} way to obtain the \ch{equations of} 
algebraic curves that \pars{\x{represent the curves that }contain}%
\x{poles\footnote{\ch{[We mention the poles for the first time!]}}}
\pars{the roots of the characteristic polynomial} regardless of the number of nodes in \ch{$\GG_{m,n}$}.
\end{itemize}

\subsection{Simple and Complex Rings}

\begin{figure}[ht] 
	\centering
	\begin{subfigure}[b]{0.24\textwidth}
		\includegraphics[width=1.1\textwidth]{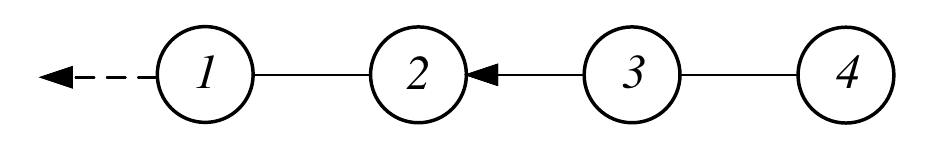}
		\caption{\ch{A} macro-vertex on 4 nodes}
		\label{fig:mv1}
	\end{subfigure}
	~ 
	\begin{subfigure}[b]{0.23\textwidth}
		\includegraphics[width=1.1\textwidth]{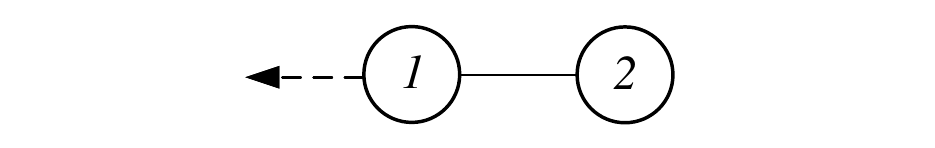}
		\caption{\ch{A} macro-vertex on 2 nodes}
		\label{fig:mv2}
	\end{subfigure}
	\caption{A \chh{macro-vertex}\x{digraph} (a) on four nodes\x{, which} can be \chh{obtained by connecting}\x{a macro-vertex of a ring digraph, is a complex consisting of} two macro-vertices of type (b) \chh{by a directed}\x{and an} arc.\x{connecting them\x{directed Hamiltonian cycle corresponding to the cyclic pursuit strategy with four agents (a) and a macro-vertex (b)}}} \label{fig:mvert}
\end{figure}

\chh{Let us find out how the set of ring digraphs is organized.}
\pars{\x{\ch{Prior to presenting} the main result, \ch{let us count non-isomorphic ring digraphs.}}
\ch{Clearly}, different macro-vertices can \ch{give rise to} \chh{isomorphic\x{the same hierarchical} ring digraphs}\x{with \ch{the} same Laplacian spectra}. For instance, consider the two macro-vertices depicted in Fig.\,\ref{fig:mvert}a,b, \ch{where each macro-vertex has an unattached dotted} arc \chh{of a Hamiltonian cycle connecting macro-vertices within a ring digraph}.\x{\chh{indicating how it}\x{es the way the macro-vertex} is connected \ch{to} its neighbor within a hierarchical \ch{di}graph.} \ch{Obviously, two macro-vertices of type (a) form the same digraph (shown in Fig.~\ref{fig_pic_0000}) as four macro-vertices of type~(b)\x{ do(?)}}.
}

\textcolor{black}{By construction, }ring digraphs\x{studied in \ch{this} paper} \ch{are scalable, i.e., they can be ``inflated'' by \chh{cloning}} \textcolor{black}{macro-vertices}. To distinguish the \ch{types of such di}graphs and characterize \ch{their simplest} components, \textcolor{black}{we introduce \chh{the following} definition.}

\begin{definition}\label{d:srd}
A ring digraph will be called a \emph{complex ring\/} \ch{if it can be represented as a Hamiltonian cycle on two or more macro-vertices. If this is not the case, we\x{will} call it a \emph{simple ring}.
A complex ring $\GG_{m,n}$ is said to be a {\em round replication\/} of a simple ring $\GG_{1,n}$ if the representations of $\GG_{m,n}$ and $\GG_{1,n}$ involve identical macro-vertices.
}
\x{...if it cannot be obtained by a round  replication\footnote{\ch{[We have not defined it yet.]}} of some other ring digraph. If a ring digraph is not a simple ring, we call it a \emph{complex ring}.}
\end{definition}

\begin{figure}[ht] 
	\centering
	\begin{subfigure}[b]{0.12\textwidth}
		\includegraphics[width=\textwidth]{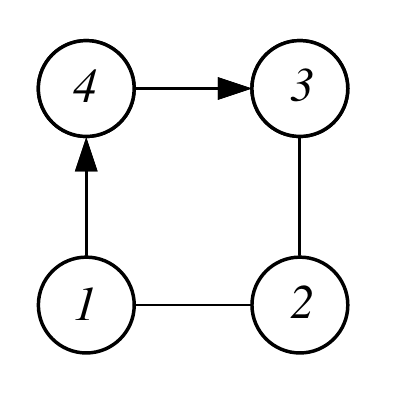}
		\caption{\ch{A} simple ring on 4 nodes}
		\label{fig:sr-1}
	\end{subfigure}
	~ 
	\begin{subfigure}[b]{0.12\textwidth}
		\includegraphics[width=\textwidth]{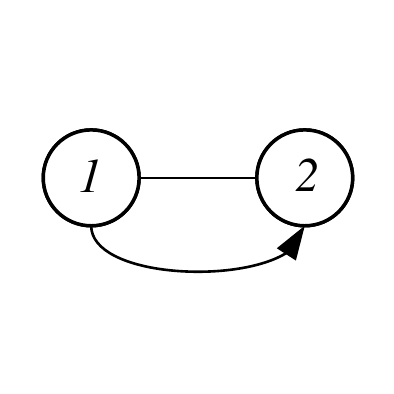}
		\caption{\ch{A} simple ring on 2 nodes}
		\label{fig:sr-2}
	\end{subfigure}
	~ 
	\begin{subfigure}[b]{0.12\textwidth}
		\includegraphics[width=\textwidth]{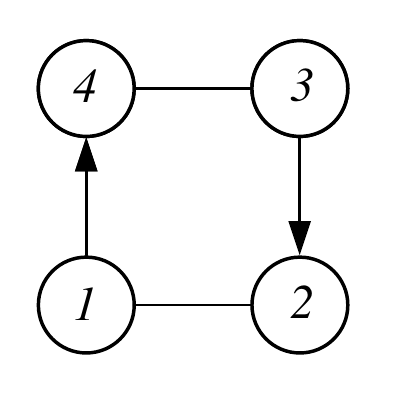}
		\caption{\ch{A} complex ring on 4 nodes}
		\label{fig:cr-1}
	\end{subfigure}
	\caption{Two simple rings \ch{((a) and (b))} and \ch{a} complex ring \ch{(c)} constructed as \ch{the} round replication of \ch{the simple ring} (b)\label{fig:dif_rings}}
\end{figure}

\chh{While} examples of simple and complex rings are \chh{shown} 
in Fig.~\ref{fig:dif_rings},\x{and} the \ch{theorem below recursively counts
} the number of non-isomorphic simple rings \ch{with a given number of nodes}.
\begin{theorem}\x{x}
\label{th:count_srd}
The number $Y(N)$ of non-isomorphic simple rings on $N$ nodes satisfies the relationship
\begin{equation}\label{eq:Y(N)}
Y(N)=\frac{2^N-\suml_{n\in D(N)}nY(n)}{N},
\end{equation}
where $D(N)$ is the set of all divisors of $N$ excluding~$N$ \ch{and $Y(1)$ is set to be~$2.$}
\end{theorem}

\proof{
\ch{First, to simplify the proof, we redefine ring digraph on $N=1$ node (\chh{cyclic pursuit of a single agent makes no sense, so this redefinition does not affect the application}) as a multidigraph that has either $1$ or $2$ directed loops. Then $Y(1)=2,$ as stated in Theorem~\ref{th:count_srd}.
Next, for any $N>1$ let us supplement the set of ring digraphs on $N$ nodes with all digraphs of the same form that additionally have arc $(N,1),$ 
where $N=mn$ (this arc is absent in  ring digraphs by definition). The supplemented set of ring digraphs will be called the set of {\em necklace digraphs}.}

\ch{Any necklace digraph on the node set $\VV=\{1\cdc N\}$ can be identified with a vector $(a_1\cdc a_N),$ where \ch{$a_i=2$ if and only if there are two opposite arcs between} nodes $i$ and $i+1 \pmod N$ and $a_i=1$ otherwise.}

\ch{A necklace digraph is {\em periodic\/} if its vector representation is periodic in the sense that $(a_1\cdc a_N)=(a_1\cdc a_n,a_1\cdc a_n\cdc a_1\cdc a_n)$ with $n<N$ being the minimum length of a subvector whose replication gives the whole vector.}\x{=modi=}

\ch{Denote by $\tY(N)$ the number of non-isomorphic non-periodic necklace digraphs on $N$ nodes.\x{=modi=}
Obviously, there is a bijection between such digraphs and distinct cycles of minimal
period~$N$ (in the case of two contractivity factors)
enumerated\footnote{Problem~3.5 ``How many different necklaces of length~$m$ can be made from beads of $q$ given
colors?'' appeared earlier in \cite{Berlekamp68}, although without the desired formula; see also~\cite{BlantonHurdMcCranie91}.} in \cite{Barnsley88} (Lemma~1 in Section~4.8). Consequently, $\tY(N)=(2^N-\sum_{n\in D(N)}n\tY(n))/N.$}
%

\ch{Finally, we prove that $Y(N)=\tY(N)$ for all $N\in\N.$ We have $Y(1)=\tY(1)$ by redefinition.
For $N>1,$ consider any non-periodic necklace digraph. Its vector representation contains at least one $a_i=1.$ Therefore, it can be transformed into the representation of a simple ring by a number of cyclic shifts transferring $a_i=1$ to the position $a_N$ corresponding to the pair of nodes $(N,1).$ This defines a one-to-one correspondence between the equivalence classes of isomorphic non-periodic necklace digraphs and the classes of isomorphic simple rings (all on $N$ nodes). Hence, the \chh{number}\x{amount} of the latter classes is given by~\eqref{eq:Y(N)}.
}
%
%
%
This completes the proof.
}

\begin{table*} 
\begin{center}
\begin{tabular}{r|rrrrrrrrrrrrrrrrrrrr}
 			   $N$	& 1& 2& 3& 4& 5& 6&  7&  8&  9& 10&  11&  12&  13&   14&   15&   16&   17&   18&     19&   20\\
\noalign{\smallskip}\hline\noalign{\smallskip}
 			$Y(N)$	& 2& 1& 2& 3& 6& 9& 18& 30& 56& 99& 186& 335& 630& 1161& 2182& 4080& 7710& 14532& 27594& 52377\\
\end{tabular}
\end{center}
 		\caption{\chh{The}\x{Several} first values of the function $Y(N)$, the number of non-isomorphic ring digraphs on $N$ nodes\label{tab:Y(N)} }
\end{table*}

\begin{corollary}\label{cor:fermat}
$1.$ If $N$ is prime, \chh{then} $Y(N)=(2^N-2)/N.$\\
$2.$ If $N=2^p,$ $p\in\N,$ then $Y(N)=(2^N-2^{N/2})/N.$
\end{corollary}

\proof{The first statement is a direct consequence of Theorem~\ref{th:count_srd}.
To prove the second one by induction, first observe that in the base case, \ch{$p=1,$ it follows from the first part}. Assume that it is \chh{true}\x{valid} for all $N=2^k,$ $k<p$ and prove it for $N=2^p.$ In this case, $D(N)=\{1,2\cdc N/2\}.$ By Theorem~\ref{th:count_srd} and the induction hypothesis, it holds that $Y(N)=(2^N-2^1-(2^2-2^1)-\ldots-(2^{N/2}-2^{N/4}))/N=(2^N-2^{N/2})/N,$ as desired.
}
\begin{figure}[ht] 
	\centerline{\includegraphics[trim={0 0cm 0 1cm},clip,width=9.5cm]{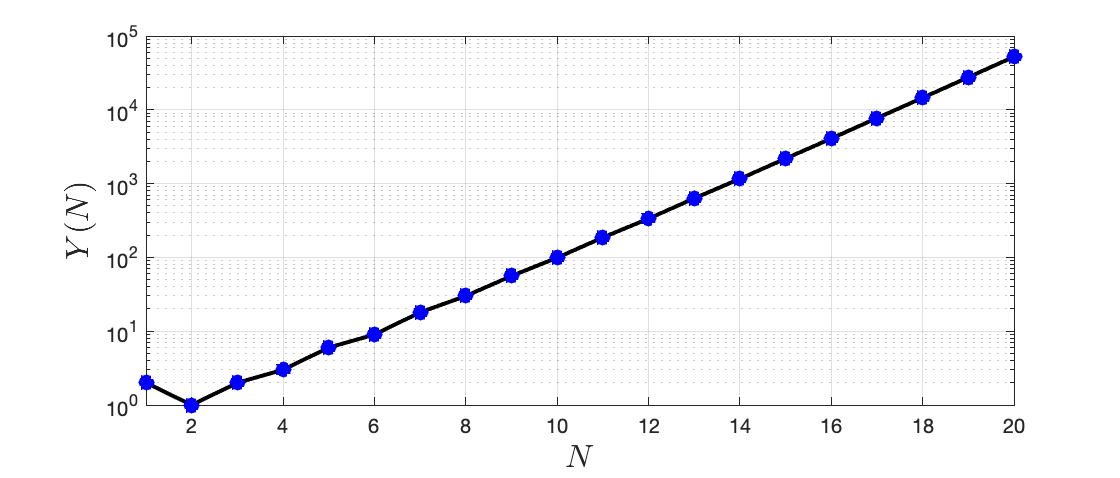}}
	\caption{The quantity $Y(N)$ as function of the number of nodes}
	\label{fig:YN}
\end{figure}

\ch{Some values of the function $Y(N)$ (modified for $N=1$) are given in Table~\ref{tab:Y(N)}.} \pars{Figure~\ref{fig:YN} illustrates \chh{its growth}\x{this table} graphically using base-10 logarithmic scale on the vertical axis.}


\begin{remark}\x{x}
\label{rem:numseq}
\ch{In the proof of Theorem~\ref{th:count_srd}, we reduced the enumeration of non-isomorphic simple rings on $N$ nodes to that of distinct cycles of minimal period~$N.$ \shch{\x{It can be noted that }Essentially the same numerical sequence 
appeared as a solution to a number of other equivalent enumeration problems including those of dimensions of the homogeneous parts of the free Lie algebras, irreducible polynomials of degree~$N$ over the field GF$(2)$, binary Lyndon words of length $N$, etc. (see sequences A001037 and A059966 in~\cite{SloanWeb}).
}}
\end{remark}

\ch{It \chh{is worth mentioning} that expression \eqref{eq:Y(N)} 
\chh{has significant consequences\x{interprets some results on the} regarding\x{in terms of} the divisibility of numbers.} Say, part~1 of Corollary~\ref{cor:fermat} implies a special case of Fermat's little theorem ($a^p \equiv a \pmod p$, where $p$ is prime) for $a=2,$ while \chh{extending \eqref{eq:Y(N)} to} multigraphs {gives a proof of this} theorem in its general form\x{ as a by-product}.}


%

\subsection{Laplacian Spectra and Algebraic Curves}
\ch{We now} consider complex rings with $N>3$ nodes and \ch{characterize the locus} of \ch{the} corresponding Laplacian spectra. 

\begin{theorem}
\label{th:high_order_curves}
\pars{For any simple ring $\GG_{1,n}$ on $n$ nodes, the Laplacian eigenvalues of all complex rings $\GG_{m,n}$ obtained by $m$-fold round replication of $\GG_{1,n}$ belong to a bounded algebraic curve of order $2n$ in \chh{$\C^+\cup \{0\}.$}}
\end{theorem}

\proof{
In accordance with Theorem~4 in \cite{AgaevChebotarev2010}, the \ch{Laplacian} characteristic polynomial of \ch{$\GG_{m,n}$ has} the form
\begin{equation}\label{eq:prod_poly}
\Delta(\la) = (P_n(\la))^{m} - (-1)^N,
\end{equation}
\ch{where} $P_n(\la) = \prod_{k=1}^{K} Z_{i_{k}}$ \ch{is an $n$th order polynomial} and $i_1, \ldots, i_K$ are the path lengths in the decomposition of the cycle $\{(1,n), (n,n - 1), \ldots , (2, 1)\}$ into
the paths linking the consecutive nodes of indegree 1 in~\chh{$\GG_{1,n}$}. The polynomials $Z_i$ are \ch{the} modified Chebyshev polynomials of the second kind:
$$
Z_n(\la) := (\la-2)Z_{n-1}(\la)-Z_{n-2}(\la),
$$
where $Z_0(\la) \equiv 1$ \ch{and} $Z_1(\la) \equiv \la - 1$.	

By\x{ virtue of} Lemma~\ref{lem.cyc}, the\x{$m$} roots \chh{$\alpha_k+j\beta_k,$ $k\in\{0\cdc m-1\}$} of \ch{$\sigma^m - (-1)^N = 0$} 
are roots of unity \chp{(the roots of $\sigma^m=-1$ are also roots of $\sigma^{2m}=1$)} lying on the unit circle in~\ch{$\C$}. Therefore, \ch{by \eqref{eq:prod_poly}, the zeros of $\Delta(\la)$ satisfy}
\begin{equation}\label{eq:P_n(lambda)}
P_n(\la) = \alpha_k+j\beta_k,\quad 
k \in \pars{\{0\cdc m-1\}},
\end{equation}
where
\begin{equation}\label{eq:circle}
\alpha_k^2 +\beta_k^2 = 1.
\end{equation}

Varying $m$ we obtain a countable set of roots of unity, which is everywhere dense on the unit circle.
\chp{This means that for any $u,v\in\R$ such that $u^2+v^2=1,$ there exist sequences $u_i\to u$ and $v_i\to v$ such that $u_i+jv_i$ are roots of unity  $(i\in\N).$
\x{\fich{Therefore, we can view the problem in context of a special case of continuous dependence of roots of polynomials on coefficients.}}
Based on this we apply Theorem~11.1 in \cite{Voev1977} on the continuous dependence of the roots of a polynomial with leading coefficient~1 on its other coefficients (cf.~\cite{Uherka1977,Hirose2020}). Due to this theorem, if $\lambda_k,$ $k\in\{0\cdc n-1\},$ are the roots of equation $P_n(\lambda)=u+jv,$ then the roots $\lambda_{k,i}$ of equations $P_n(\lambda)=u_i+jv_i$ $(k\in\{0\cdc n-1\}, i\in\N)$ can be numbered in such a way that $\lambda_{k,i}\to\lambda_k,$ $k\in\{0\cdc n-1\}.$
This justifies the following method for determining the curve (in the implicit form $f(x,y)=0$) on which the Laplacian eigenvalues of complex rings $\GG_{m,n}$ are everywhere dense.
Setting $\la=x+jy$ for \eqref{eq:P_n(lambda)} and substituting ${\rm Re}[P_n(x+jy)]=\alpha_k$ and ${\rm Im}[P_n(x+jy)]=\beta_k$ into \eqref{eq:circle} yields an equation of order $2n,$ which determines the desired algebraic curve of order~$2n$ in the form $f(x,y)=0.$ Indeed, this curve contains the roots of \eqref{eq:P_n(lambda)} for all $\alpha_k+j\beta_k$ that belong to the unit circle. According to the above continuity theorem, any neighborhood of each such a root contains infinitely many roots of  \eqref{eq:P_n(lambda)} in which $\alpha_k+j\beta_k$ are roots of unity. The latter roots lie on the same curve and are the Laplacian eigenvalues of ring digraphs~$\GG_{m,n}.$}
\x{We can obtain the zeros of $\Delta(\la)$ in the implicit form $f(x,y)=0,$ where $x$ and $y$ are the coordinates in $\R^2,$
by substituting $\la=x+jy$ into \eqref{eq:P_n(lambda)} and equating the real and imaginary parts in both sides taking into account~\eqref{eq:circle}.%
\x{This leads to an equation \chh{of the form} $f(x,y)=0,$ where $f(x,y)$ is an order $2n$ polynomial.
Its zero set is an algebraic curve of order~$2n.$}
\textcolor{black}{\chh{In more detail,}\x{Indeed, the $N$ roots of polynomial~\eqref{eq:prod_poly} can be found from \eqref{eq:P_n(lambda)}.}
substituting ${\rm Re}[P_n(x+jy)]=\alpha_k$ \chh{and} ${\rm Im}[P_n(x+jy)]=\beta_k$ into \eqref{eq:circle} we arrive at \chh{an equation of order~$2n.$}}
\textcolor{black}{Note that equation \eqref{eq:circle} is insensitive to the signs of\x{the variables} $\alpha_k$ and $\beta_k$. However, this does not lead to the appearance of \x{``extra''}\chh{any parts of the resulting curve}\x{elements} that do not contain the \chh{eigenvalues of}\x{spectrum of the Laplacian matrix} $\mathcal{L}_N$ \chh{with varying~$N$. Indeed, if $m$ is even, then}\x{The explanation for this fact is as follows: for even values of $m$,} the roots of (negative) unity\x{\footnote{\chh{[We do not have negative unity as $N$ is even.]}}} are located symmetrically about both axes in~$\C$. That is, changing the signs of \chh{$\alpha_k$ and/or $\beta_k$}\x{the real/imaginary part} will not allow to go beyond the original set of roots. \x{This is not the case }\chh{If $m$ is odd, then the set of roots $\{\alpha_k+j\beta_k,\,k\in\{0\cdc m-1\}\}$ of the equation $\sigma^m = 1$ or $\sigma^m = -1$ is symmetric about the real axis only, while changing the sign of $\alpha_k$ transforms any root of the first equation into a root of the second one and vice versa. However, all of them are the roots of $\sigma^{2m} = 1,$ which also appear in the right-hand side of some equations~\eqref{eq:P_n(lambda)}. This means that odd values of $m$ do not bring ``extra'' roots either.\x{eigenvalues of the matrices $\mathcal{L}_N$}}\x{The roots of (negative) unity do not have this property for odd~$m$. However, it is easy to show that the following condition is true:
the set of roots of $\sigma^{2m} = 1$
contains the roots of $\sigma^{m} = -1$, see Fig.~\ref{fig:rtunit}.
Obviously roots $\sigma_l$, $l \in \{0\cdc 2m-1\}$, of the first equation equal the roots $\sigma_k$, $k \in \{0\cdc m-1\}$, of the second one for $l = 2k+1$.\x{ As mentioned above, for an even $m$, the sign change does not affect the set of roots.}}
}
}

\ch{\x{Since the entire set of roots of unity is everywhere dense on the unit circle, the corresponding zeros of $\Delta(\la),$ \chh{which depend on the coefficients of \eqref{eq:P_n(lambda)} continuously,} are everywhere dense on the resulting algebraic curve. \chh{Thus, this curve} contains the Laplacian spectra locus of \chh{the set of}\x{all} complex rings $\GG_{m,n}$ obtained \chh{from a given simple ring  $\GG_{1,n}$} by\x{$m$-fold} replication.}
By the properties of the Laplacian spectra of digraphs, \chp{they lie}\x{this locus lies} in \chh{$\C^+\cup \{0\}$}.
\chh{Substituting $\la=|\la|(\cos\varphi+j\sin\varphi)$ into $P_n(\la)=\la^n+\sum_{k=0}^{n-1}p_k\la^k$ for $\la\ne0$ we have $|P_n(\la)|=|\la|^n\,|1+\sum_{k=0}^{n-1}p_k\la^{-n+k}(\cos{k\varphi}+j\sin{k\varphi})|.$ Therefore, it is easy\x{straightforward} to specify $h>0$ such that $|\la|>h$ implies $|P_n(\la)|>1.$ Consequently, $\la$ with $|\la|>h$ cannot satisfy \eqref{eq:P_n(lambda)} and thus the Laplacian spectra locus of ring digraphs $\GG_{m,n}$ is bounded.}
This completes the proof.}
}

\chh{Let us emphasize} that an unbounded ``inflation'' of a \chh{ring digraph $\GG_{m,n}$ by increasing}\x{due to the growth of} $m$ leaves the Laplacian eigenvalues on the same algebraic curve and only increases their density on it.

\begin{corollary}
\chh{For a fixed $n\in\N$, }\ch{the number of distinct algebraic curves of order $2n$ 
containing the Laplacian spectra of ring digraphs obtained by round replication of simple rings on $n$ nodes does not exceed the number of non-isomorphic simple rings on $n$ nodes determined by Theorem~\ref{th:count_srd}.}
\end{corollary}

\subsection{Quartic and Sextic Curves}
\textcolor{black}{In this section, we consider several special cases that allow\x{ for} relatively simple closed-form expressions of the \chh{corresponding}\x{respective} algebraic curves mentioned in \chh{Theorem~\ref{th:high_order_curves}}\x{the previous section}.}

\begin{figure}[ht] 
	\centerline{\includegraphics[width=6.375cm]{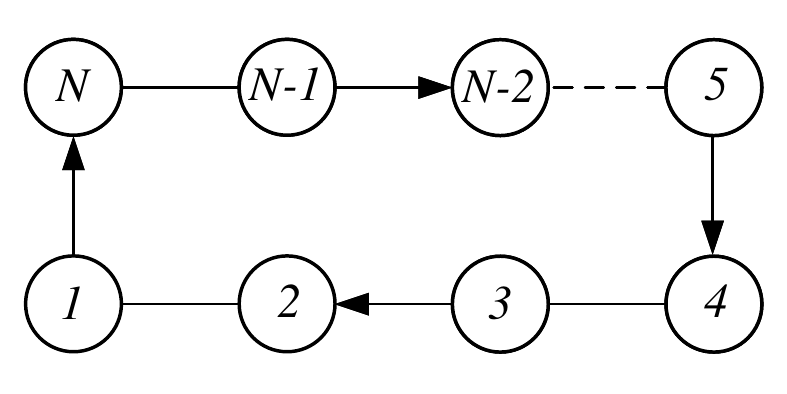}}
	\caption{\ch{A} round replication of the \ch{simple ring shown} in Fig.~\ref{fig:sr-2}}
	\label{fig_pic0}
\end{figure}

\emph{\textcolor{black}{\underline{The case $n=2$.}}} We first\x{,} consider a complex ring with \ch{the following}\x{a specific} structure:
It has $N = 2m$ nodes, $m \geq 2$, and contains a Hamiltonian cycle supplemented \chh{by}\x{with} the inverse cycle, where every second arc is dropped\x{ out} \ch{(\x{see }Fig.\,\ref{fig_pic0})}. This \ch{di}graph is a \ch{round} replication of the simple ring depicted in Fig.~\ref{fig:sr-2}\chh{; the ring digraph in Fig.~\ref{fig_pic_0000} belongs to this class with $m=4.$}


The Laplacian matrix 
\ch{of this digraph \chh{has}\x{ is of} the form}
\begin{equation}
\label{eq:laplacian0}
\LL_{N} =
\begin{bmatrix}[r]
2 & -1 & 0 & 0 & \cdots & 0 & -1\\
-1 & 1 & 0 & 0 & \cdots & 0 & 0 \\
0 & -1 & 2 & -1 & \cdots & 0 & 0  \\
\vdots & \vdots & \ddots & \ddots & \ddots & \vdots & \vdots \\
0 & \cdots & 0 & -1 & 1 & 0 & 0 \\
0 & \cdots & 0 & 0 & -1 & 2 & -1  \\
0 & \cdots & 0 & 0 & 0 & -1 & 1
\end{bmatrix},
\end{equation}
\ch{and by \eqref{eq:prod_poly}, its} characteristic polynomial \ch{is} $(Z_2)^{\frac{N}{2}}-1$ \textcolor{black}{$ = (\la^2-3\la+1)^m-1$. Its roots satisfy
$$
\la^2-3\la+1-\alpha_k-j\beta_k = 0, \quad k \in \{0\cdc m-1\}.
$$
From $(x+jy)^2-3(x+jy)+1-\alpha_k-j\beta_k = 0$ it follows $\alpha_k = (x-1.5)^2-y^2-1.25$ and $\beta_k = 2xy-3y$. Substituting the last expressions into formula~\eqref{eq:circle} gives the equation of the curve.}

\begin{figure}[ht] 
	\centerline{\includegraphics[trim={0 7cm 0 7cm},clip,width=9.5cm]{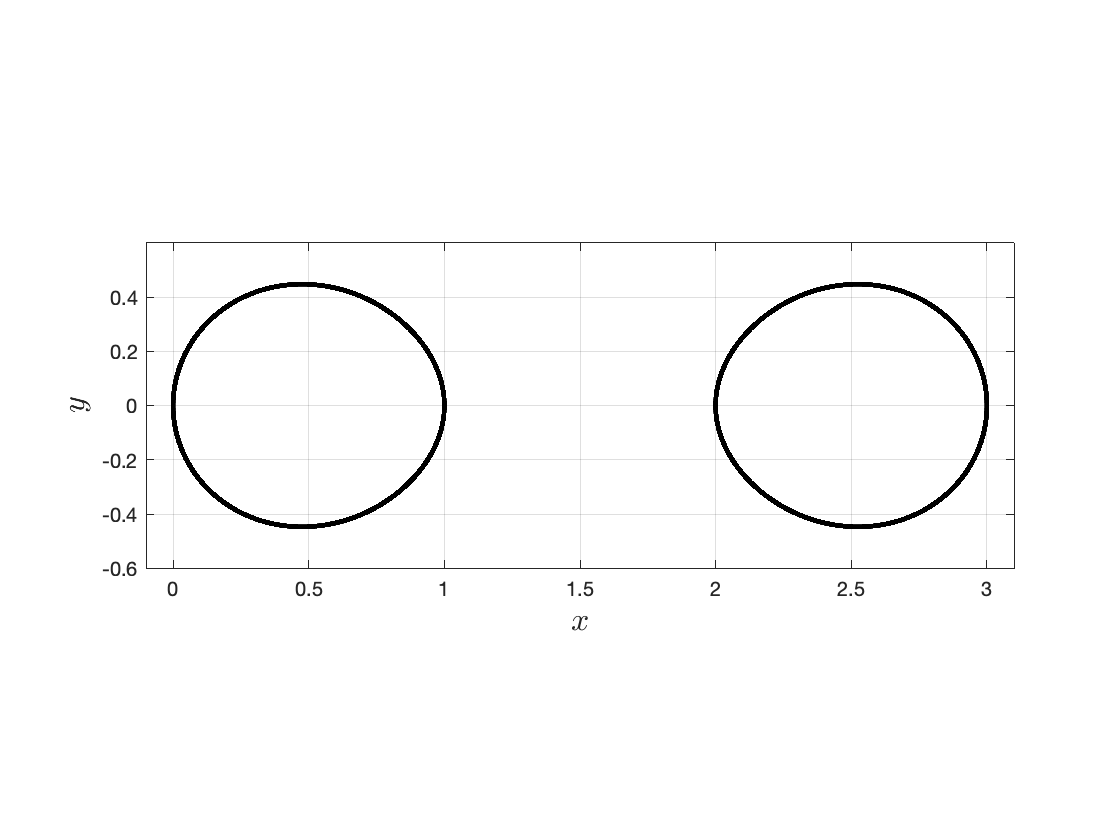}} 
	\caption{Cassini ovals}
	\label{fig:cassini}
\end{figure}

In this case, the eigenvalues of \ch{the} Laplacian matrix \eqref{eq:laplacian0} lie on \ch{the} quartic Cassini curve 
(\x{or}Cassini ovals) defined by
\begin{equation}
\label{eq:cassini_ovals}
	[(\tilde x-\sqrt{5})^2+\tilde y^2][(\tilde x+\sqrt{5})^2+\tilde y^2] = 2^4,
\end{equation}
where $\tilde x = 2(x-3/2)$ \ch{and} $\tilde y = 2y$, see \cite{ParChe2018} for the details.
\ch{This} curve is \ch{shown} in Fig.~\ref{fig:cassini}.

\begin{figure}[ht] 
	\centering
	\begin{subfigure}[b]{0.14\textwidth}
		\includegraphics[width=\textwidth]{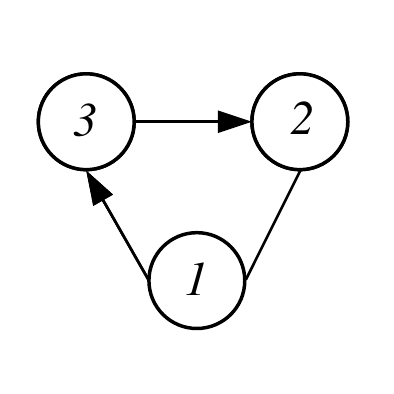}
		\caption{Simple ring \ch{\#1}}
		\label{fig:sr_3-1}
	\end{subfigure}
	\begin{subfigure}[b]{0.14\textwidth}
		\includegraphics[width=\textwidth]{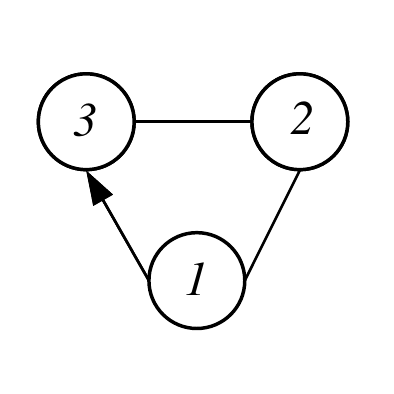}
		\caption{Simple ring \#2}
		\label{fig:sr_3-2}
	\end{subfigure}
	\caption{Two simple rings on $n=3$ nodes}\label{fig:simple_rings_3}
\end{figure}

\begin{figure}[ht] 
	\centerline{\includegraphics[width=8.5cm]{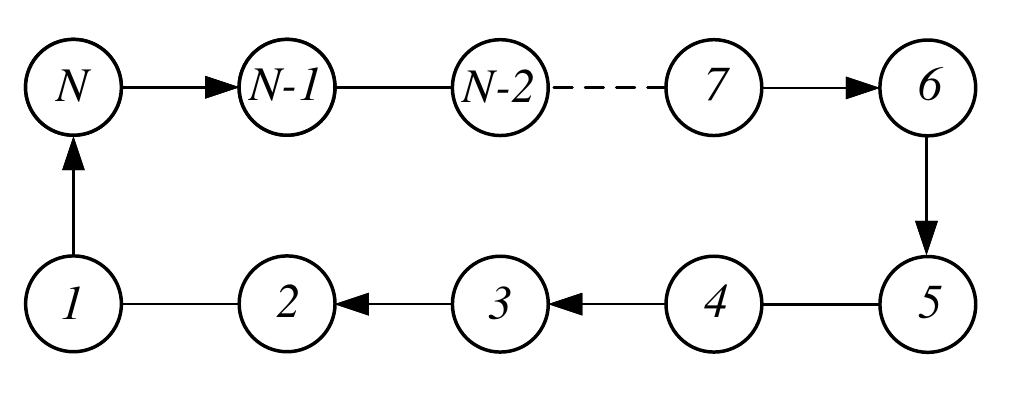}}
	\caption{\ch{The} \chh{ring digraph} obtained by round replication of the \ch{simple ring} in Fig.~\ref{fig:sr_3-1}}
	\label{fig:pic1}
\end{figure}

\emph{\textcolor{black}{\underline{The case $n=3$.}}} \ch{\x{We }Observe} that there are \ch{exactly} two \ch{non-isomorphic} simple rings on $n=3$
nodes\ch{; these are} depicted in Fig.~\ref{fig:simple_rings_3}.
\ch{Consider} two complex rings on $N = 3m$ nodes \ch{($m > 1$)} \ch{constructed by round} replication of \ch{these} simple rings. \ch{The one\x{\ch{in} Fig.~\ref{fig:sr_3-1}} obtained from simple ring~\#1 is} shown
in Fig.~\ref{fig:pic1}.

Its Laplacian matrix \ch{has} the form
\begin{equation}
\label{eq:laplacian1}
\LL_{N} =
\begin{bmatrix}[r]
	2 & -1 & 0 & 0 & \cdots & 0 & -1\\
	-1 & 1 & 0 & 0 & \cdots & 0 & 0 \\
	0 & -1 & 1 & 0 & \cdots & 0 & 0  \\
	\vdots & \vdots & \ddots & \ddots & \ddots & \vdots & \vdots \\
	0 & \cdots & 0 & -1 & 2 & -1 & 0 \\
	0 & \cdots & 0 & 0 & -1 & 1 & 0  \\
	0 & \cdots & 0 & 0 & 0 & -1 & 1
\end{bmatrix},
\end{equation}
\ch{and by \eqref{eq:prod_poly}, its} characteristic polynomial \ch{is} $(Z_1 Z_2)^\ch{m} - (-1)^N$.

\begin{figure}[ht] 
	\centerline{\includegraphics[trim={0 4cm 0 5cm},clip,width=9.5cm]{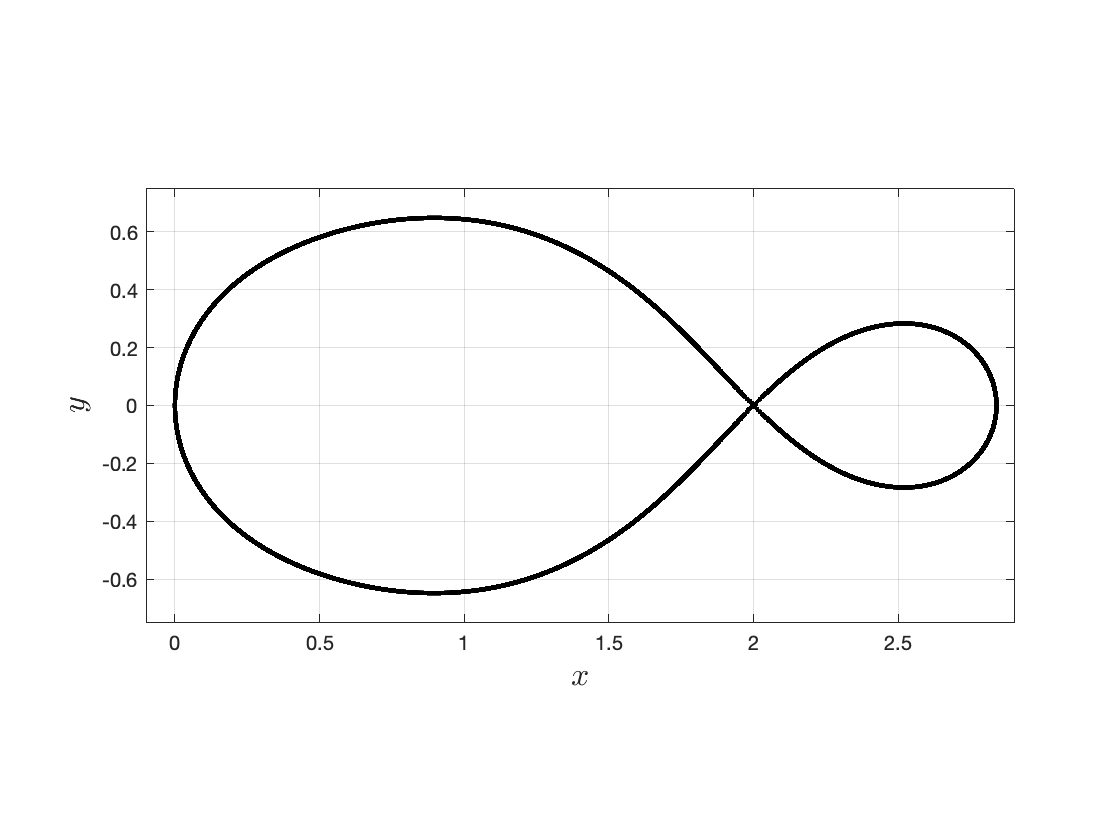}}
	\caption{The sextic curve defined by \eqref{eq:sexticcurve1}}
	\label{fig_pic4}
\end{figure}

According to Theorem~\ref{th:high_order_curves}, the eigenvalues of \ch{the matrix} \eqref{eq:laplacian1} lie on \ch{a} sextic curve. \ch{Its} equation \ch{is}
\begin{eqnarray}\label{eq:sexticcurve1}
(\tilde x^2+\tilde y^2)^3+(4+4 \tilde x)(\tilde x^2+\tilde y^2)^2-2 \tilde{x}^3-4 \tilde x^2\\
+\,6 \tilde x \tilde y^2+4 \tilde y^2 = 0,\nonumber
\end{eqnarray}
where $\tilde x = x-2$ \ch{and} $\tilde y = y$.
%
\ch{This} curve\x{\eqref{eq:sexticcurve1}} is depicted in Fig.~\ref{fig_pic4}.

\begin{figure}[ht] 
	\centerline{\includegraphics[width=8.5cm]{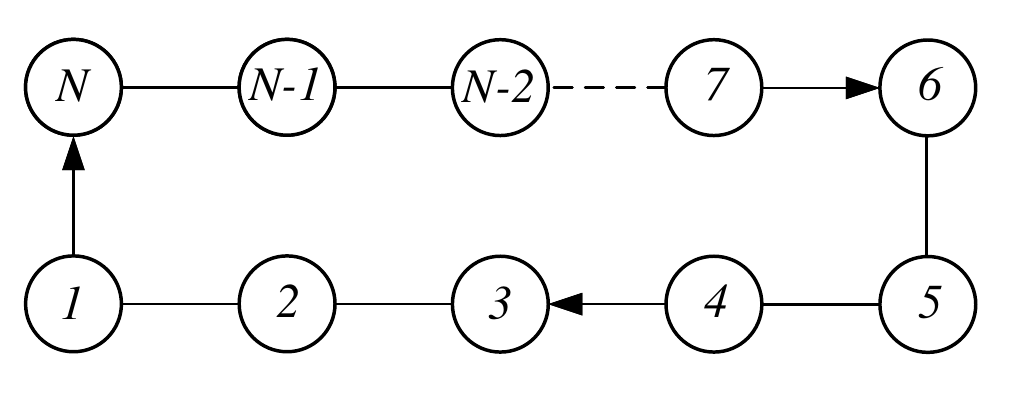}}
	\caption{\ch{The} \chh{ring digraph} obtained by round replication of the \ch{simple ring} in Fig.~\ref{fig:sr_3-2}}
	\label{fig:pic2}
\end{figure}

\ch{The} complex ring constructed \ch{by round} replication of simple ring~\#2 (Fig.~\ref{fig:sr_3-2}) \ch{is shown in} Fig.~\ref{fig:pic2}.

\ch{Its Laplacian matrix is of the form}
\begin{equation}
\label{eq:laplacian2}
\LL_{N} =
\begin{bmatrix}[r]
2 & -1 & 0 & 0 & \cdots & 0 & -1\\
-1 & 2 & -1 & 0 & \cdots & 0 & 0 \\
0 & -1 & 1 & 0 & \cdots & 0 & 0  \\
\vdots & \vdots & \ddots & \ddots & \ddots & \vdots & \vdots \\
0 & \cdots & 0 & -1 & 2 & -1 & 0 \\
0 & \cdots & 0 & 0 & -1 & 2 & -1  \\
0 & \cdots & 0 & 0 & 0 & -1 & 1
\end{bmatrix};
\end{equation}
\ch{and by \eqref{eq:prod_poly}, its characteristic polynomial is} $(Z_3)^\ch{m}-(-1)^N$.

\begin{figure}[ht] 
	\centerline{\includegraphics[trim={0 7cm 0 7cm},clip,width=9.5cm]{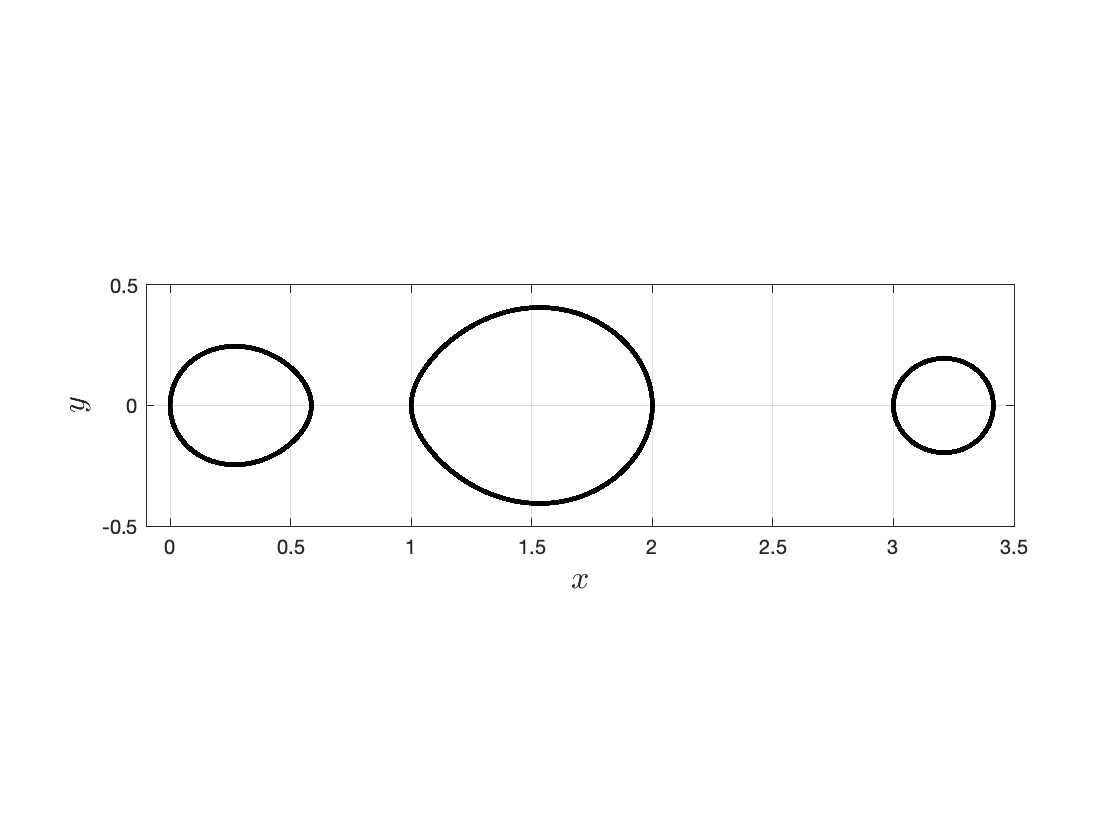}}
	\caption{The sextic curve defined by \eqref{eq:sexticcurve2} }
	\label{fig_pic33}
\end{figure}

\ch{By Theorem~\ref{th:high_order_curves}, the} eigenvalues of \ch{the} matrix \eqref{eq:laplacian2} lie on \ch{a} sextic curve; \ch{it is} defined by \chh{equation}
\begin{eqnarray}\label{eq:sexticcurve2}
(\tilde x^2+\tilde y^2)^3+2 \tilde x(\tilde x^2+\tilde y^2)^2-3 \tilde{x}^4-6\tilde x^3+2 \tilde x^2 \tilde y^2\\
+\,2 \tilde{x}^2+ 2 \tilde{x} \tilde y^2+4 \tilde{x} +5\tilde{y}^4+6\tilde{y}^2= 0,\nonumber
\end{eqnarray}
where $\tilde x = x-2$ \ch{and} $\tilde y = y$.
%
\ch{This} curve\x{defined by \eqref{eq:sexticcurve2}} is depicted in Fig.~\ref{fig_pic33}.

Graphs with a more complex structure \ch{based on} simple rings on \ch{$4,5,\ldots$} nodes can be obtained in the same way\x{ as shown above,} along with \ch{the} corresponding expressions for higher-order curves \pars{that contain the spectrum loci.}

\textcolor{black}{In subsection~\ref{ss_D}, we \chh{present a} result \chh{involving} a weighted \ch{necklace}\x{ring} digraph. \chh{Such a} structure generalizes the \chh{topology} of cyclic pursuit in a\x{ somewhat} different way: \chh{There are no macro-vertices, but the arcs of one of the directions are weighted and have the same weight}.} \pars{Due to the presence of \chh{this variable weight}, the corresponding Laplacian spectra \chh{belong to a certain drop-shaped \emph{region} rather than lie on an algebraic curve.}}

\subsection{A Weighted Ring}
\label{ss_D}
%

\x{We}Consider a \ch{\emph{weighted necklace}}\x{ring} digraph on \ch{$N$ nodes consisting of} a Hamiltonian cycle and the inverse one.

Assume \chh{that} all arcs of one of the cycles have the same weight~$a$, and the arcs in the opposite direction have weight~$b$. Without loss of generality we can \chh{restrict ourselves to}\x{ analyze} the case where one weight is\x{ equal to} unity and the other one is $c\in [0,~1]$.

\begin{figure}[ht] 
	\centerline{\includegraphics[width=3.5cm]{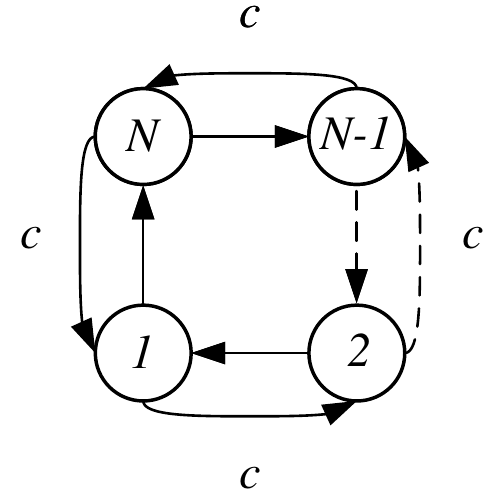}}
	\caption{\chh{A}\x{The} two-cycle weighted\x{ring} digraph}
	\label{fig:weighted_ring-1}
\end{figure}

\chh{A}\x{The} \ch{di}graph \ch{of this type is shown} in Fig.~\ref{fig:weighted_ring-1}.

Its Laplacian matrix \chh{has} the form
\begin{equation}
\label{eq:laplacian_weight-1}
\LL_N =
\begin{bmatrix}[r]
\!1+c\! & \!-c\! & \!0\! & \!0\! &\! \cdots\! & \!0\! & \!-1\!\\
\!-1\! & \!1+c\! & \!-c\! & \!0\! &\! \cdots\! & \!0\! & \!0\! \\
\!0\! & \!-1\! & \!1+c\! & \!-c\! &\! \cdots\! & \!0\! & \!0\!  \\
\!\vdots\! & \!\vdots\! & \!\ddots\! & \!\ddots\! &\!\ddots\! &\!\vdots\! & \!\vdots\! \\
\!0\! & \!\cdots\! & \!0\! & \!-1\! & \!1+c\! & \!-c\! & \!0\! \\
\!0\! & \!\cdots\! & \!0\! & \!0\! & \!-1\! & \!1+c\! & \!-c\!  \\
\!-c\! & \!\cdots\! & \!0\!& \!0\! & \!0\! & \!-1\! & \!1+c\!
\end{bmatrix}.
\end{equation}
\begin{lemma}
\label{lem:ellipses}
For any weight \ch{$c\in [0,~1]$} and any \ch{$N\in\N,$} the eigenvalues of \ch{matrix} \eqref{eq:laplacian_weight-1} lie on the ellipse
\begin{equation}\label{e:elli}
\frac{(x-(1+c))^2}{(1+c)^2}+\frac{y^2}{(1-c)^2} = 1.
\end{equation}
\end{lemma}

\begin{figure}[ht] 
	\centerline{\includegraphics[trim={0 4cm 0 5cm},clip,width=9.5cm]{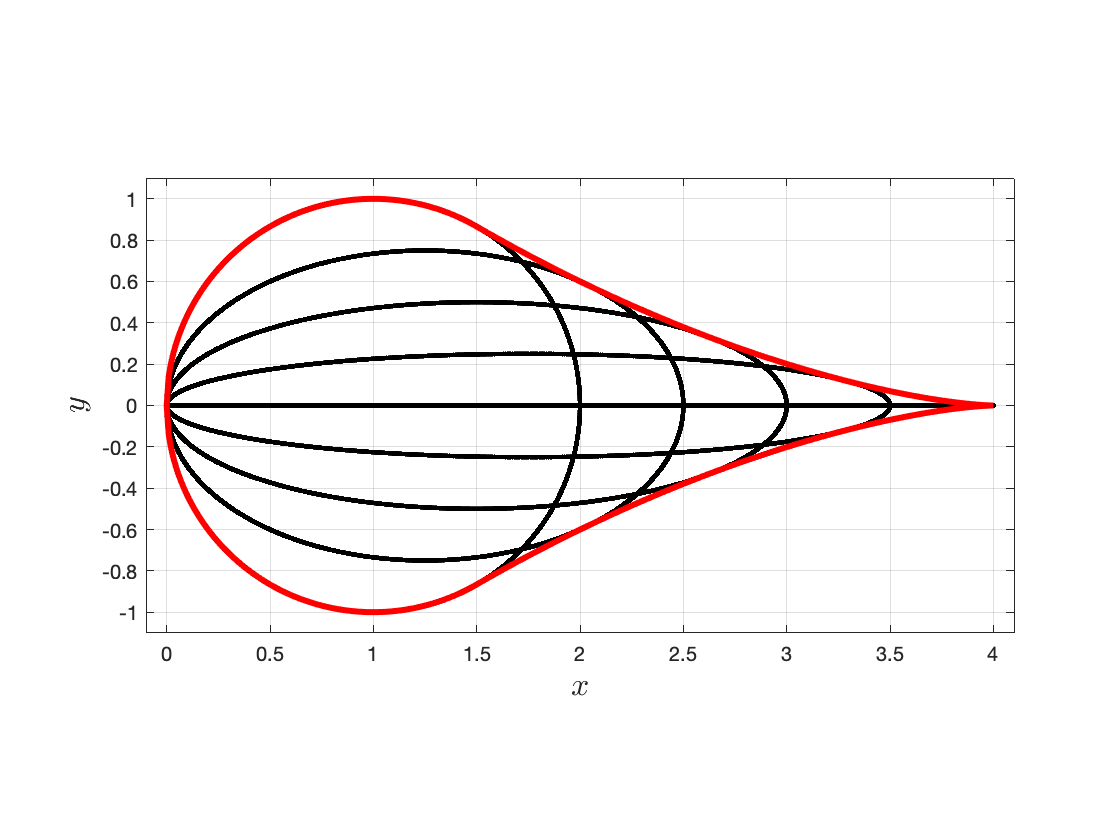}}
	\caption{\pars{A sequence of five ellipses \pars{that contain} \ch{the spectrum loci of the Laplacian matrices \eqref{eq:laplacian_weight-1}} as $c$ increases from $0$ to~$1,$ including \chh{a unit} circle $(c=0)$ and \chh{a}\x{the} segment $(c=1)$; the \ch{boundary $f_{1,2}(x)$ of a} drop-shaped region, \chh{which is the union of}\x{ containing} all the ellipses \ch{(see Theorem~\ref{th:ellipses}), is shown in red}}}
	\label{fig:seq_ellipses}
\end{figure}

\proof{
\ch{Obviously,} 
$\LL_N = (1+c)I_N - \mathcal{P}_N - c\mathcal{P}_N^{N-1}$, where $\mathcal{P}_N$ is \ch{the counter-clockwise principal circulant permutation 
matrix}~\eqref{eq:perm_cycl_purs}. \ch{Therefore, the eigenvalues} of the Laplacian \ch{matrix are}\x{obey}
$\la_k = (1+c)-\e^{j \frac{2\pi k}{N}}-\ch{c\e^{j\frac{2\pi(N-1)k}{N}}}$, \ch{$k\in \pars{\{1\cdc N\}}$}. Rewriting \ch{this expression} in \ch{a} trigonometric form leads to \ch{the} parametric equation of \ch{the} ellipse \ch{\eqref{e:elli} in $\R^2$, which completes} the proof.
}
\begin{remark}
\label{rm:ellipses}
\pars{The limit cases of \eqref{e:elli} are the unit circle centered at $(1,\,0)$ $($for $c=0)$ and the segment $[0,\,4]$ of the real axis $($for $c=1)$. \chh{These along with three ellipses of the form~\eqref{e:elli}} are shown in Fig.~\ref{fig:seq_ellipses}.}
\end{remark}

\begin{theorem}
\label{th:ellipses}
Every eigenvalue of \ch{matrix} \eqref{eq:laplacian_weight-1} for any $c \in [0,~1]$ and $\ch{N\in\N}$ \ch{lies in the} drop-shaped region \ch{bounded by the functions}
\begin{eqnarray}
\nonumber
\lefteqn{\ch{f_{1,2}(x)}\! =\!}\\
&\label{eq:dropp}\!\!\!\!\!\!\begin{cases}
\pm \sqrt{1-(x-1)^2}                                                        &\!\!\mbox{if } x \!\in\! \ch{[0,1.5]};\\
\pm \frac{1}{\sqrt{2}}(3\!-\!\sqrt{1+2x})\sqrt{\sqrt{1\!+\!2x}\!-\!x\!+\!1} &\!\!\mbox{if } x \!\in\! (1.5,4].
\end{cases}
\end{eqnarray}
\end{theorem}

\vspace{ .1in}
\proof{
\ch{For} ellipses \eqref{e:elli}
, we have $x\in[0, 2(1+c)]$ and $y\in[-(1-c),(1-c)],$ with the maximum and minimum at $x=1+c$ (cf.\ Fig.~\ref{fig:seq_ellipses}). Thus, for any two different ellipses \ch{of this family}, each one extends beyond the other. Let us \chh{fix}\x{an arc weight} $c\in(0,1).$\x{\pars{[$\,]0,1[\,$ or $(0,1)$ ???? \ch{$]0,1[$ is the Bourbaki notation less common in the US.}]}}
Suppose that $(x_{cz},\pm y_{cz})$  with $x_{cz}\ne0$ are the intersection points of the two ellipses corresponding to arc weights $c$ and $z\ne c.$ Then $x_{cz}$ increases in~$z.$ Let $f_z(x)$ be the function representing the upper (non-negative) part of the ellipse corresponding to~$z\in(0,1).$ We have
\begin{eqnarray}
\label{e:maj}
&f_z(x)\!>\!f_c(x)\mbox{\;whenever\;}\\
&\nonumber ((z\!<\!c)\,\&\,(0\!<\!x\!<\!x_{cz}))
\mbox{\,or\,}  
               ((c\!<\!z)\,\&\,(x_{cz}\!<x\!\le\!2+2c)).\nonumber
\end{eqnarray}
Let
$$
x_c=\liml_{z'\to c-0, \,z''\to c+0}x_{z'z''}=\liml_{z'\to c-0}x_{z'c}=\liml_{z''\to c+0}x_{cz''}.
$$

It follows from \eqref{e:maj} that the only $x$ for which $f_c(x)=\maxl_zf_z(x)$ is~$x_c.$

Let us find $x_c$ as a function of~$c$. To this end, we \ch{first} find $x_{cz}$ as a function of $c$\x{!!!} and~$z.$ Using \eqref{e:elli} it is straightforward to verify that

\begin{equation}
x_{cz}=2\frac{\frac{(1-z)^2}{1+z}-\frac{(1-c)^2}{1+c}}{\left(\frac{1-z}{1+z}\right)^2-\left(\frac{1-c}{1+c}\right)^2}.
\end{equation}

\ch{Now} it \chh{can be shown} that
\begin{equation}\label{e:x_c}
x_c=\lim_{z\to c}x_{cz}=\frac{(1+c)(3+c)}2,
\end{equation}
and by \eqref{e:elli} it holds that
\begin{equation}\label{e:f_cx_c}
f_c(x_c)=\frac12(1-c)\sqrt{(1-c)(3+c)}.
\end{equation}

Substitution of the expression for $c$ from Eq.\,\eqref{e:x_c} into \eqref{e:f_cx_c} yields the form of $f_{1,2}(x)$ given in Theorem~\ref{th:ellipses}.
}

In the following section, we show how the 
localization \ch{of the} Laplacian spectra helps to analyze \ch{the} stability of networks of high-order agents.

\section{A Consensus Criterion}\label{sec_consens}

\subsection{The Consensus Region}

\chh{A}\x{The} system composed of agents \eqref{eq:gen_cons_system0} \chh{controlled}\x{governed} by distributed\x{ control} \chh{protocol} \eqref{eq:gen_cons_system01} can be equivalently represented as
\begin{equation}
\label{eq:gen_cons_system}
{\rm \bf a}(s)x = {\rm \bf b}(s)(-\LL_Nx),
\end{equation}
where $s :=  \frac{d}{dt}$,  $x = [x_1,~x_2,\ldots, x_N]^{\top}$\ch{, and} $\LL_N$ is the Laplacian matrix of the \chh{dependency} digraph $\GG_N$ containing a spanning converging tree.

\x{Consider a network system with\x{the} dynamics \ch{given by}\x{obeying equation}~\eqref{eq:gen_cons_system}.}
The following condition simplifies \ch{the} analysis of reaching \ch{consensus} in system~\eqref{eq:gen_cons_system} by \chh{dividing}\x{splitting} the problem into two subproblems.

\begin{definition}[{\rm\hspace{-4pt} \cite{PolyakTsypkin1996, HaraTanakaIwasaki2014, LiDuan2017}}]
\label{def:consensusregion}
The consensus \ch{region (or $\Omega$-region)} of the function  $\phi(s) = {\rm \bf a}(s)/{\rm \bf b}(s)$
in the Laplace variable $s$ is the set of points $\la$ \ch{in} $\C$ for which the function $\phi(s) -\la$ has no zeros in the closed right
half-plane:
$$\Omega = \{\la \in \C: \phi(s)-\la \neq 0~ \text{\rm whenever } {\rm Re}(s) \ge 0\}.$$
\end{definition}
\ch{\medskip}
The function $\phi(s)$ is sometimes referred to as the \emph{generalized frequency variable}~\cite{HaraHayakawaetal2007},~\cite{HaraTanakaIwasaki2014}.

\begin{lemma}[{\rm\hspace{-3pt} \cite{PolyakTsypkin1996, HaraTanakaIwasaki2014, LiDuan2017}}]
\label{lem:consensusregion}
The network system with agents described by \eqref{eq:gen_cons_system0} reaches consensus under protocol \eqref{eq:gen_cons_system01} if and only if
$$
\la_i \in \Omega,\quad i \in \pars{\{2\cdc N\}},
$$
where $\la_i$, $i \in \pars{\{2\cdc N\}},$ are the nonzero eigenvalues of $-\LL_N$.
\end{lemma}

The details of determining the consensus region may be found in \cite{PolyakTsypkin1996}. \pars{In the case of $\phi(s) = s^2+\gamma s$, $\gamma > 0,$ this region has \ch{the} form of the interior of a parabola in the complex plane: $\phi(j\omega) = -\omega^2+j\gamma \omega$, $-\infty < \omega < \infty$, and if $\phi(s) = s$, \chh{then} the $\Omega$-region is the open left half-plane of the complex plane.}


\subsection{\x{Application}Consensus in Systems on Ring Digraphs}

In this \ch{sub}section, we formulate \ch{and prove} a consensus criterion \chh{for systems~\eqref{eq:gen_cons_system}}.
\begin{theorem}
\label{th:cons_suff}
\chh{A} system \eqref{eq:gen_cons_system}, where $\LL_N$ is the Laplacian matrix of \ch{a} 
ring \ch{dependency} digraph$,$ reaches consensus in the sense of \eqref{eq:consen_def} for all\x{possible} number\ch{s} of agents\x{$N>2$[?]} if and only if the \ch{locus of the spectrum of $-\LL_N$} lies entirely
in \ch{the} \pars{open} consensus region $\Omega$ defined by $\phi(s)$ \ch{and shares only the point} $(0,\,j0)$ with it\ch{s boundary}.\x{\pars{share one common point ????? or just forget about this point, it's more or less evident, since we care only about nonzero eigenvalues?} \ch{[I think, accurately mention this point every time will be the clearest way]}}
\end{theorem}
\proof{
\pars{\ch{By Theorem~\ref{th:high_order_curves}, \x{In the paper, we construct and analyze special graphs with ring structure, such that the corresponding} the Laplacian spectra of ring digraphs $\GG_{m,n}$ obtained by round $m$-fold replication from\x{of} a given simple ring \chh{$\GG_{1,n}$}\x{on~$n$ nodes} lie on a certain} algebraic curve of order $2n$, irrespective of~\ch{$m$}.
\x{This means, whatever number of nodes in graph we have, either small or infinitely large, the curve will remain the same.} Taking this fact into account, \ch{it suffices to apply} Lemma~\ref{lem:consensusregion} to prove Theorem~\ref{th:cons_suff}.}
}
\chq{\begin{remark}
\label{rm:order1}
As mentioned above,  Theorem~\ref{th:cons_suff} applies to systems whose ring topology always contains a spanning converging tree, which guarantees\x{ attaining} consensus in the case of first-order agents. Thus, this theorem gives additional conditions that ensure consensus at a higher order of agents.
\end{remark}
}

\subsection{Consensus in\x{ Specific} Networks of Second-Order Agents}
Consensus problems in networks of second-order agents \ch{have been widely studied; \chh{see, e.g.,}} \cite{RenCao2011, Goldin2013, Bullo2018, Ren2008a}. Here we \ch{consider}\x{illustrate} the cases with absolute and relative velocity \pars{gain} from the point of view of the consensus criterion of Theorem~\ref{th:cons_suff}.
\ch{Thus,} the consensus conditions derived for the examples below are based\x{either} on finding\x{the straightforward solution of the equation corresponding to} the intersection of\x{the boundary of}\x{curve that encircles} the consensus region and \pars{the curve that contains the} \chh{spectrum of\x{ the} system matrix~\chh{ $-\LL_N$}}\x{Laplacian \ch{spectra} locus}. \ch{In some cases, we will use}\x{with possible application of} Vieta's theorem\x{to the same equation with \ch{an} analysis of \ch{the} signs of its terms}.

\begin{example}
\label{exmpl:parabola}
Consider \ch{the following} system of $N$ interconnected second-order agents with absolute velocity \pars{gain} $\gamma > 0$ (\pars{see \cite{ParChe2018} for the details}):
\begin{equation}\label{e:ex1}
\ddot x +\gamma \dot x = -r \LL_Nx,
\end{equation}
where $r > 0$ is a scaling \chh{factor. This factor}\x{coefficient, parameter} \pars{is introduced for the sake of generality and can be considered either as\x{a} part of agent's dynamics or \ch{as} a parameter of the \ch{communication} Laplacian matrix.} \chh{In any case, matrix $-r\LL_N$ now plays the role of $-\LL_N$ in\x{terms of} Theorem~\ref{th:cons_suff}.}

The consensus region \chh{of system~\eqref{e:ex1}} 
is bounded by the curve\x{defined by the function} $\phi(j\omega) = -\omega^2+j\gamma \omega$, and the corresponding curve in $\R^2$ has the form $y^2 = -\gamma^2 x$.
\ch{By Theorem~\ref{th:cons_suff}}, the \ch{system reaches consensus}\x{condition \ch{reduces} to the condition that} \pars{if and only if} the spectrum of $-r\LL_N$
belongs to the interior of the parabola $ y^2 = -\gamma^2x$ (except for the intersection at the origin) for all~$N$.

Consider the\x{case of} communication topology represented by a Hamiltonian cycle (\chh{the classical cyclic pursuit illustrated by Fig.~\ref{fig:ham}}) \ch{as the dependency digraph}. The corresponding Laplacian matrix is given by \eqref{eq:laplacian_cycl_purs}; therefore, the eigenvalues of $-r \LL_N$ are located on the circle \ch{of radius $r$} centered at $(-r,~j0)$. \ch{It is straightforward to check that this circle has no intersection with the above parabola except for the origin point whenever%
}
\textcolor{black}{$r/\gamma^2 \leq 1/2$.} \fich{Note that this result for the ``predecessor–follower'' topology corresponds to the condition of asymptotic stability of the platoon solution  in \cite[Theorem~2]{RoggeAeyels2008}, as $N$ tends to infinity.}

If the \ch{dependency} digraph has the form shown in \chh{Fig.~\ref{fig_pic0}}\x{\ref{fig:cassini}}, then the system reaches consensus in the sense of \eqref{eq:consen_def} if and only if the\x{corresponding} Cassini ovals \eqref{eq:cassini_ovals} \chh{(Fig.~\ref{fig:cassini}) reflected about the vertical axis and $r$-scaled} belong to the consensus region.
This is satisfied \ch{whenever}
\textcolor{black}{$r/\gamma^2 \leq {7}/{6}$.} \fich{In terms of the vehicular platoon control problem, this result means that the system becomes eventually unstable \x{if}\chq{when} the above inequality does not hold.}

\pars{Figures~\ref{fig:omegacircle_parab} and \ref{fig:omegacassini_parab} \chh{illustrate} the cases \ch{where} the condition of Theorem~\ref{th:cons_suff} is\x{either} satisfied or violated\x{, respectively}.}

\begin{figure}[ht] 
  \centering
    \includegraphics[trim={0 4cm 0 5cm},clip,width=9.5cm]{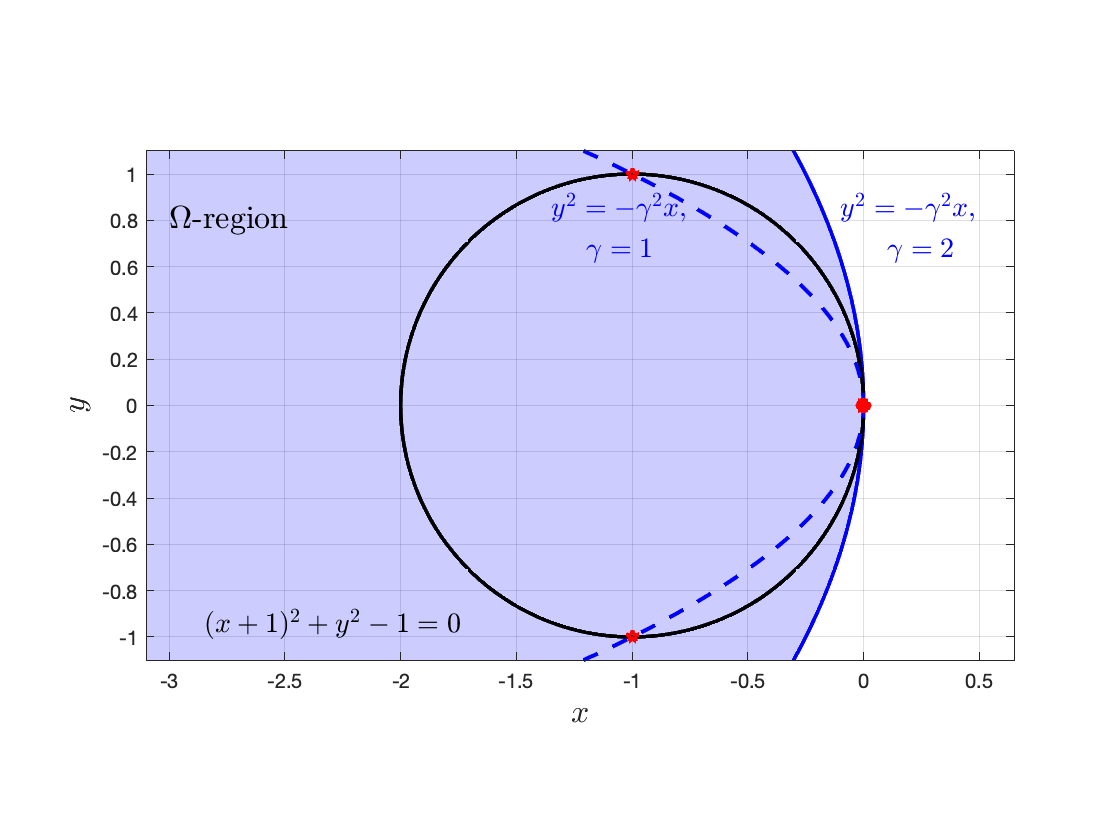}
    \caption{The $\Omega$-region bounded by $y^2 = -\gamma^2 x$ and the unit circle ($r=1$), \chh{where} $\gamma \in \{1,~ 2\}$}
  \label{fig:omegacircle_parab}
\end{figure}
\end{example}

\begin{figure}[ht] 
  \centering
    \includegraphics[trim={0 4.8cm 0 5.8cm},clip,width=9.5cm]{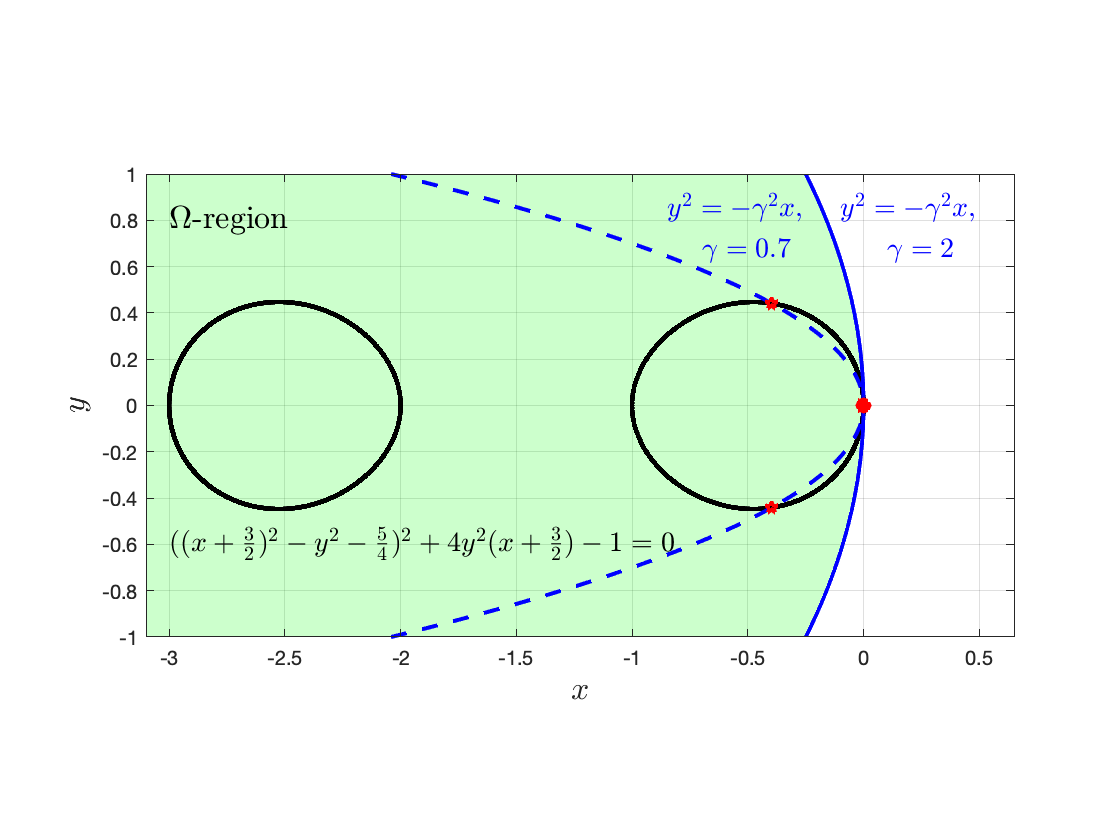}
    \caption{The $\Omega$-region bounded by $y^2 = -\gamma^2 x$ and the \chh{reflected Cassini ovals \eqref{eq:cassini_ovals} ($r = 1$)}, where $\gamma \in \{0.7,~2\}$}
  \label{fig:omegacassini_parab}
\end{figure}

\begin{example}
\label{exmpl:relvel}
Now consider the system
\begin{equation}
\label{e:ex2}
\ddot x = -r\LL_Nx -\gamma r \LL_N \dot x, \quad r>0
\end{equation}
with relative velocity \pars{gain} $\gamma > 0$ and $r > 0$.

Here\x{,} the generalized frequency variable is $\phi(s) = s^2/(1+\gamma s)$. Since $\phi(j\omega) = -\omega^2/(1+\gamma^2\omega^2)+j\gamma \omega^3/(1+\gamma^2\omega^2)$, the \ch{boundary of the consensus region \chh{of system~\eqref{e:ex2}} on} the plane $\R^2$ \chh{has algebraic expression} $y^2 = -\gamma^2 x^3/(\gamma^2x+1)$.

\ch{Similarly to} the previous example, consider two \chh{communication topologies and the two corresponding}\x{versions of the} \ch{curves containing the} spectrum of $-r \LL_N$\chh{:} \ch{(i)~}the circle \ch{of radius} $r$ centered at $(-r,~j0)$ and \ch{(ii)~}the Cassini ovals \eqref{eq:cassini_ovals} \chh{reflected about the vertical axis and $r$-scaled}. In the first case, there always exists an intersection at $x = -2r/(1+2r\gamma^2)$. \ch{In the second case,} the corresponding cubic equation always has one negative real root \ch{$x_0$ regardless of} the values of $r$ and~$\gamma,$ \ch{as illustrated in} \pars{Figs.~\ref{fig:omegacircle_instab} and \ref{fig:omegacassini_instab}.}
\begin{figure}[ht] 
  \centering
    \includegraphics[trim={0 5.8cm 0 6.5cm},clip,width=9.5cm]{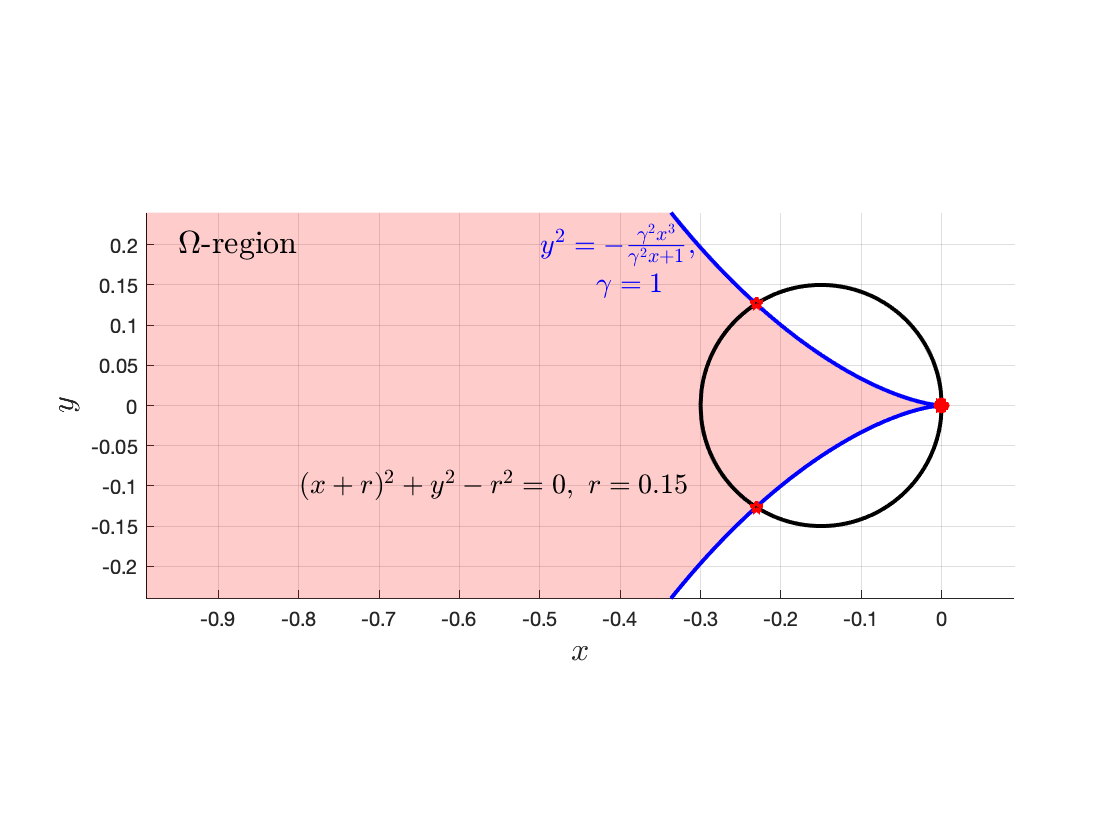}
    \caption{The $\Omega$-region bounded by $y^2 = -\frac{\gamma^2 x^3}{\gamma^2x+1}$ and the circle \chh{containing the spectrum of $-r\LL_N$\x{Laplacian spectra locus}, where}  $\gamma = 1$ and $r = 0.15$}
  \label{fig:omegacircle_instab}
\end{figure}
\begin{figure}[ht] 
  \centering
    \includegraphics[trim={0 5.8cm 0 6.5cm},clip,width=9.5cm]{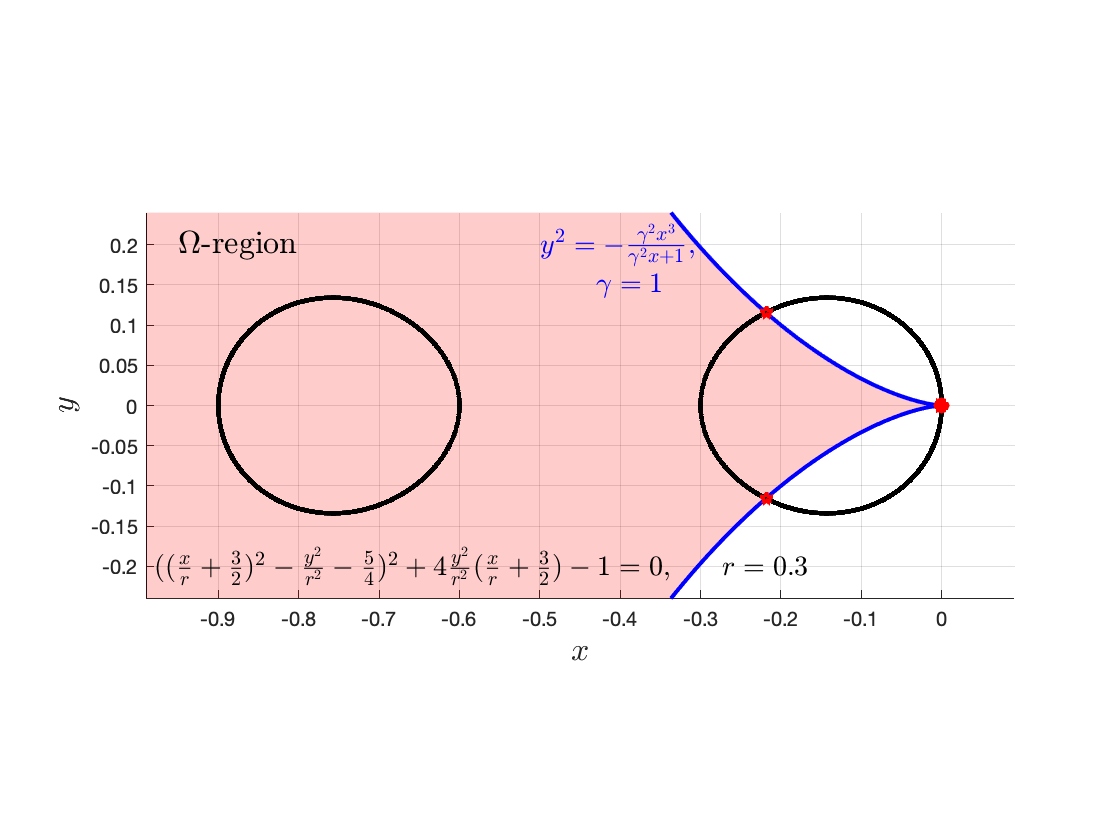}
    \caption{The $\Omega$-region bounded by $y^2 = -\frac{\gamma^2 x^3}{\gamma^2x+1}$ and the Cassini ovals\x{\eqref{eq:cassini_ovals}} \chh{containing the spectrum of $-r\LL_N$, where} $\gamma = 1$ and $r = 0.3$}
  \label{fig:omegacassini_instab}
\end{figure}
\end{example}

\begin{corollary}\label{co:nonetopo}
\chh{For system \eqref{e:ex2} with predefined relative velocity \pars{gain} $\gamma$,} \ch{no} \x{of second order agents}\chh{cyclic topology\x{considered in \ch{this} paper with the corresponding} whose Laplacian spectrum belongs to the} curve \eqref{eq:uncirc}, \eqref{eq:cassini_ovals}, \eqref{eq:sexticcurve1}, \eqref{eq:sexticcurve2}, \chh{or \eqref{eq:dropp} guarantees} consensus for \pars{all $N\in\N$}. \fich{For vehicle platoons control problems, this means that the system is eventually unstable.}\x{[The footnote is not needed as the formulation seems correct enough.]\footnote{\pars{It would be more correct to say that here we consider the curves reflected about the vertical axis, as we discuss the spectra of $-r\LL_N$.}}}
\end{corollary}

{\it Sketch of the proof:}\;
\textcolor{black}{\chh{Observe} that both the curve $y^2 = -\gamma^2 x^3/(\gamma^2x+1)$ bounding the consensus domain \chh{of system \eqref{e:ex2}} and the\x{other} curve containing the Laplacian spectrum \chh{of $-r\LL_N$} share the \chh{ origin} point $(0,~j0)$.
Near \chh{this point}, under a negative increment of~$x$, \chh{the positive branch of} any of the curves \chh{under consideration\x{listed above that} containing} the Laplacian \chh{spectra of $-r\LL_N$ grows faster than that of} the curve $y^2 = -\gamma^2 x^3/(\gamma^2x+1),$ \chh{which can be straightforwardly confirmed by} the analysis of\x{their} derivatives.\x{The derivations are straightforward but bulky and hence are not presented.}
Therefore, starting from the origin,\x{it is clear that} \chh{all the positive branches of} the \chh{spectra} curves\x{that contain the spectra always} lie above the \chh{positive branch} of the boundary curve. \chh{Thus}\x{As it follows from here}, they do not belong to the $\Omega$-region. \chh{Consequently, by Theorem~\ref{th:cons_suff}, none of the topologies listed in Corollary~\ref{co:nonetopo} guarantees  consensus for all $N\in\N.$}}

\begin{remark}
\label{rm:relvel}
\ch{\x{Obviously,}It \chh{follows} \textcolor{black}{from the analysis of the spectrum \chh{of} $-r\LL_N$} that}
\chh{system~\eqref{e:ex2}} with\x{a specified in advance} \ch{a certain value of the} relative velocity \pars{gain} $\gamma$ can reach \ch{consensus in the sense of \eqref{eq:consen_def}, provided that the} number of agents $N$ \ch{is sufficiently small}. For example, for $\gamma = 3.4$, the system with uni-directed topology reaches consensus \pars{if and only if}
$N \leq 6$.\x{However, for the number $N = 7$ the consensus condition is violated.} \ch{With a slightly increased} \chh{factor}\x{coefficient} $\gamma = 4$, the \ch{system always reaches consensus \pars{if and only if $N \leq 7$}}, see Fig.~\ref{fig:omegacircle_eigs}\x{condition \ch{is no longer violated}}.
\begin{figure}[ht] 
  \centering
    \includegraphics[trim={0 5.8cm 0 6.5cm},clip,width=9.5cm]{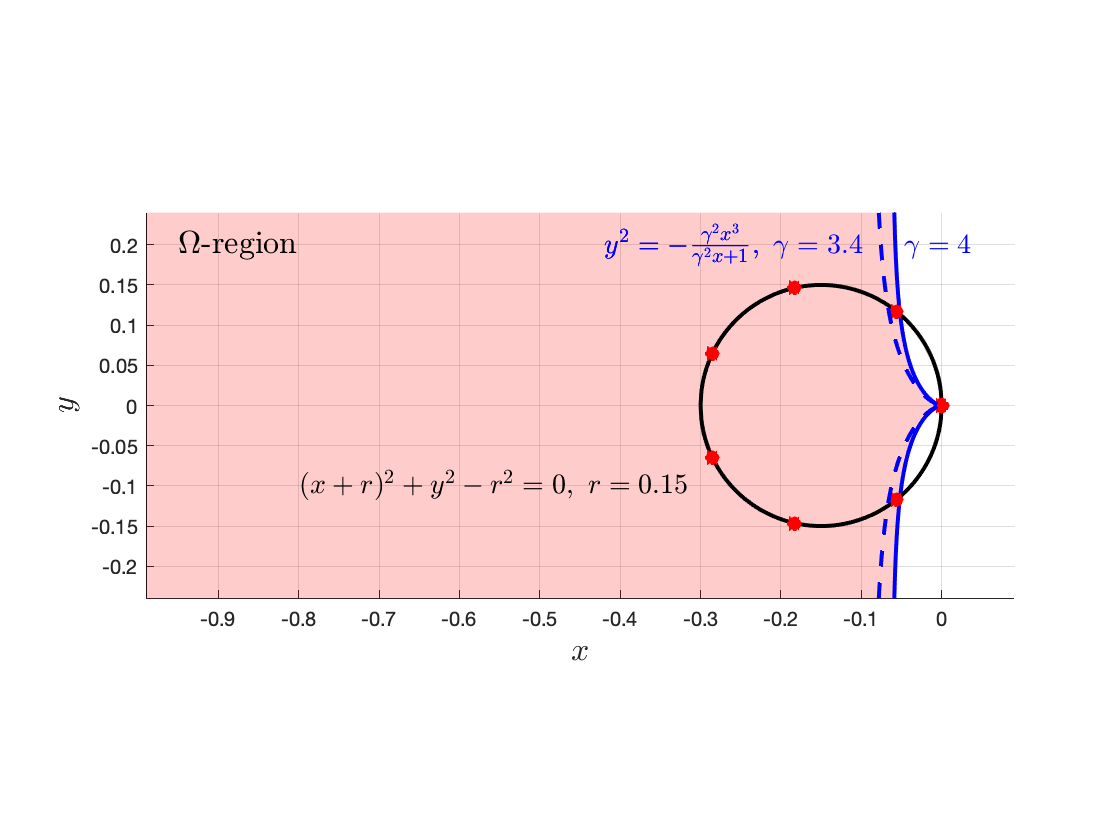}
    \caption{The $\Omega$-region bounded by $y^2 = -\frac{\gamma^2 x^3}{\gamma^2x+1}$, the \pars{circle that contains the\x{Laplacian} spectra locus \chh{of $-r\LL_N$}}\x{scaled unit circle} ($r = 0.15$, $\gamma \in \{3.4,~4$\}), and the eigenvalues of the matrix  \chh{for} $N=7$}
  \label{fig:omegacircle_eigs}
\end{figure}
\end{remark}

\begin{example}
\label{exmpl:hara}
Let the system \chh{have}\x{satisfy} the dynamics \x{\ch{[Here $\LL_N$ appears with + when $r>0$, which diverges velocities. What is the physics of this dynamics and does it interfere with consensus?]}} \x{\pars{[wrong sign, $x$ should be with a positive coefficient, not $\dot x$]}} \x{\ch{It seems that now, if the group was stationary, but in different positions, then it starts to diverge and remains doing that slowly, but even with acceleration, isn't it?}}
$$
\ddot x = -\frac{r}{\gamma}\LL_Nx +\left(r \LL_N -\frac{1}{\gamma}I_N\right)\dot x, \quad \gamma, r>0\x{,}
$$
and\x{\ch{[Here it looks like an independent assumption(?)]} \pars{[Should be: with a more exotic]}}\x{have} \ch{a} more exotic generalized frequency variable\x{is} $\phi(s) = (s+\gamma s^2)/(1-\gamma s)$\pars{\cite{HaraTanakaIwasaki2014}}. For $s = j\omega$ we have $\phi(j\omega) = -2\gamma^2\omega^2/(1+\gamma^2\omega^2)+j\gamma (\omega - \gamma^2\omega^3)/(1+\gamma^2\omega^2)$, with the \ch{boundary of the consensus region $\Omega$} in $\R^2$ expressed as $y^2 = -x(1+\gamma x)^2/(\gamma(\ch{2+\gamma x}))$.

\x{Now}Consider a uni-directed topology, \ch{whose Laplacian spectrum lies on a} circle. It can be shown \ch{that} the consensus condition of Theorem~\ref{th:cons_suff} is satisfied if and only if 
\textcolor{black}{$r\gamma \leq 0.25$}. \pars{The consensus region and two \ch{versions} of \pars{the circle that contains the\x{Laplacian} spectra locus \chh{of $-r\LL_N$}} are depicted in Fig.~\ref{fig:omegacircle_exotic}.} \ch{Consensus is reached for $r=0.15,$ but this is not the case with $r=0.35$.}
\begin{figure}[ht] 
  \centering
    \includegraphics[trim={0 4.6cm 0 5.3cm},clip,width=9.5cm]{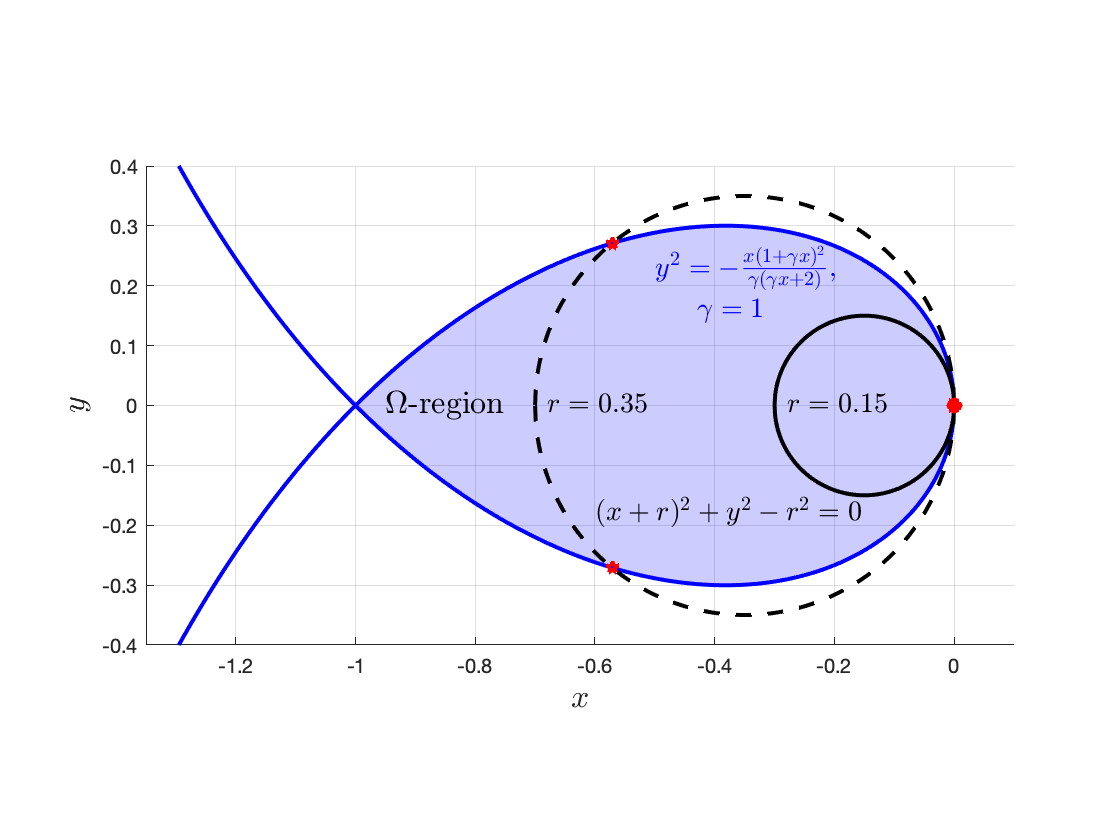}
    \caption{The $\Omega$-region bounded by $y^2 = -x(1+\gamma x)^2/(\gamma(\gamma x+2))$ and the \pars{circles that contain the\x{Laplacian} spectra loci \chh{of $-r\LL_N$}}, where $r \in \{0.15,~0.35\}$ \chh{and} $\gamma = 1$}
  \label{fig:omegacircle_exotic}
\end{figure}
\end{example}

\section{Conclusions}

Cyclic pursuit is one of the most attractive and interesting problems of network\x{ed} communication. \ch{Its properties are studied using its Laplacian spectrum, which allows} for \pars{exact} localization on \ch{the} unit circle. In \ch{this} paper, we studied \chh{several} \ch{versions of {\em hierarchical}} cyclic pursuit, where each macro-vertex \ch{of the dependency digraph} is a sequence of directed and \ch{bi-directional} arcs.

The contribution of the paper is threefold. For the network dynamical systems on ring digraphs\ch{,} we\ch{:}
\begin{itemize}
    \item proved that the \ch{corresponding} Laplacian spectra \ch{ lie on certain} high-order algebraic curves regardless
of the number of macro-vertices in the network\ch{;}
    \item \x{showed\chno{,} that} presented \ch{an algorithm for obtaining implicit equations of} \ch{these} curves;
    \item proposed a consensus condition in \ch{the} frequency domain\x{, regardless of the} \ch{applicable to any} number of agents in the network.
\end{itemize}

\chh{A characteristic feature of the algebraic curves obtained in this study is that they contain the spectrum loci of specific (Laplacian) matrices associated with network dynamical systems. Some of them, such as the Cassini ovals, have a simple geometric interpretation~\cite{Lawrence1972}; some others do not seem to have appeared in handbooks on special functions.}
\x{\textcolor{magenta}{We also note that the algebraic curves discussed above are seemingly not presented in
the corresponding books ...CITE SMTH(?)...; they do not result from the ``typical'' geometric considerations,
but rather have the meaning of spectrum loci of specific matrices arising in network systems.}}

Possible extensions \chh{of this work} include\x{Laplacian} spectra localization of more general\x{types of} weighted \chh{networks}\x{digraphs} \ch{that represent hierarchical pursuit}\x{alternative ways of introducing hierarchy}. These problems \ch{are} the subject of continuing research.

 \section{Acknowledgements}
 Research of P. Shcherbakov in Section IV was
supported by the Russian Science Foundation (project No. 21-71-30005).

\begin{IEEEbiography}[{\includegraphics[width=1in,height=1.25in,clip,keepaspectratio]{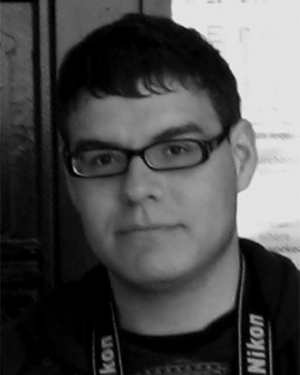}}]{Sergei E. Parsegov } received the M.S. degree in automation and control from Bauman Moscow State Technical University, Moscow, Russia, in 2008 and the Ph.D. (Candidate of Science) degree in physics and mathematics from the Institute of Control Sciences, Russian Academy of Sciences, in 2013. He is a Senior researcher in the Laboratory of Robust and Adaptive Systems, Russian Academy of Sciences. From 2018 he is with the Center for Energy Science and Technology of Skolkovo Institute of Science and Technology (Skoltech), Moscow. His research interests are dynamics and control of complex networks, opinion dynamics in social networks, distributed optimization and control in energy systems.
\end{IEEEbiography}

\begin{IEEEbiography}[{\includegraphics[width=1in,height=1.25in,clip,keepaspectratio]{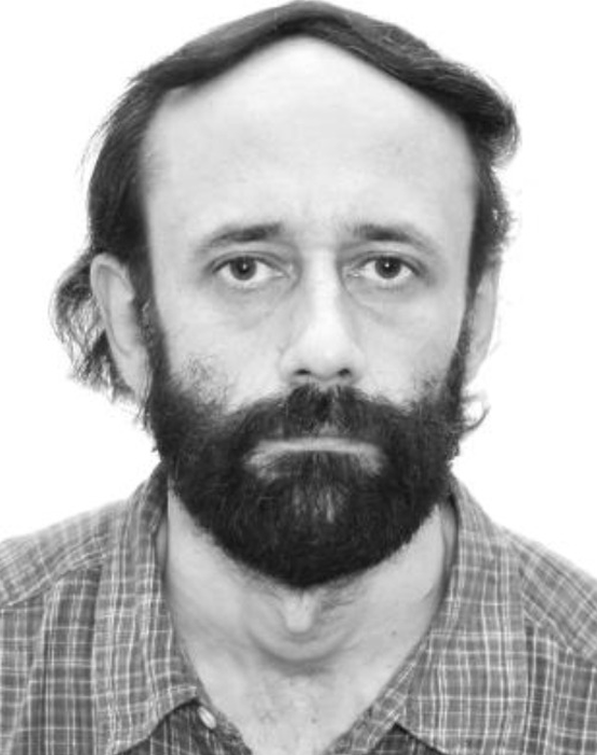}}]{Pavel Yu. Chebotarev} received the Ph.D. (Candidate of Science, 1990) and Doctor of Science (2008) degrees in physics and mathematics from the Institute of Control Sciences, Russian Academy of Sciences. He is the head of the Laboratory of Mathematical Methods for the Analysis of Multi-agent Systems. His research interests are in algebraic graph theory, clustering, decentralized control, voting theory, and social dynamics.
\end{IEEEbiography}

\begin{IEEEbiography}[{\includegraphics[width=1in,height=1.25in,clip,keepaspectratio]{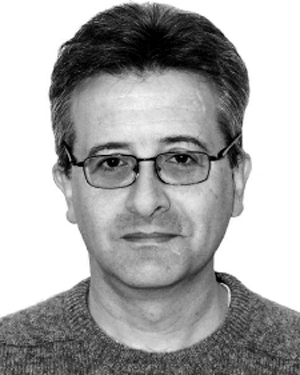}}]{Pavel S. Shcherbakov}  was born in Moscow, Russia, in 1958. He received the M.S. degree in applied mathematics from the Department of Applied Mathematics, Moscow University of Transportation, in 1980, and the Candidate of Science degree (Russian equivalent of Ph.D.)
and the Doctor of Science Degree both (in physics and mathematics) from the Institute of Control Science, Russian Academy of Sciences, Moscow, in 1991 and 2004, respectively. Since 1988, he has been with the Department of Robust and Adaptive Systems (Tsypkin Lab), Institute of Control Science, Moscow, where he is currently a Principal Researcher. His interests are in parametric robustness of control systems, probabilistic and randomized methods in control, linear matrix inequalities, and invariant ellipsoid methods in systems and control theory.
\end{IEEEbiography}

\begin{IEEEbiography}[{\includegraphics[width=1in,height=1.25in,clip,keepaspectratio]{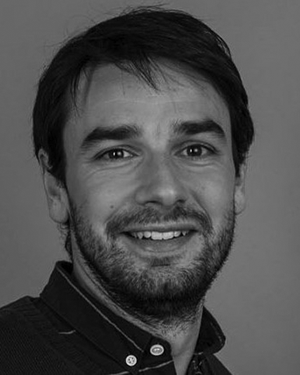}}]{Federico Mart\'in Ib\'a\~nez} was born in Buenos Aires, Argentina, in 1982. He received the B.S. degree in electronic engineering from National Technological University (UTN), Buenos Aires, in 2008, and the Ph.D. degree in power electronics from the University of Navarra, San Sebastian, Spain, in 2012. From 2006 to 2009, he was with the Electronics Department, UTN. From 2009 to 2016, he was with the Power Electronics Group, Centro de Estudios e Investigaciones Tecnicas de Gipuzkoa. He is currently an Assistant Professor with the Center for Energy Science and Technology, Skoltech, Moscow, Russia. His research interests are in the areas of high-power dc–dc and dc–ac converters for applications related to energy storage, supercapacitors, electric vehicles, and smartgrids.
\end{IEEEbiography}

\end{document}